\documentclass[10pt,journal,compsoc]{IEEEtran}

\usepackage{cite}
\usepackage{graphicx, wrapfig}
\usepackage{footmisc}
\usepackage{amssymb}
\usepackage{bm}
\usepackage{times}
\usepackage{helvet}
\usepackage{courier}
\usepackage{times}
\usepackage{mathrsfs}
\usepackage{amsmath}
\usepackage{booktabs}
\usepackage{algorithm}
\usepackage{algorithmicx}
\usepackage{algpseudocode}
\usepackage{booktabs}
\usepackage{color}
\usepackage{epsfig}
\usepackage{balance}
\usepackage{eucal}
\usepackage{rotating}
\usepackage{multirow}
\usepackage{diagbox}
\usepackage{xspace}
\usepackage{ragged2e}
\usepackage{bbding}
\usepackage[tight,footnotesize]{subfigure}

\newtheorem{myDef}{Definition}
\newtheorem{exmp}{Example}

\newcommand{\ie}{\emph{i.e.,}\xspace}
\newcommand{\aka}{\emph{a.k.a.,}\xspace}
\newcommand{\eg}{\emph{e.g.,}\xspace}

\newcommand{\ignore}[1]{}

\begin{document}

% The file aaai.sty is the style file for AAAI Press
% proceedings, working notes, and technical reports.
%
\title{Heterogeneous Information Network Embedding for Recommendation}
\author{Chuan~Shi,~\IEEEmembership{Member,~IEEE}, Binbin~Hu, Wayne~Xin~Zhao~\IEEEmembership{Member,~IEEE}\\ and Philip~S.~Yu,~\IEEEmembership{Fellow, IEEE}}

\IEEEtitleabstractindextext{
\begin{abstract}
\justifying
%With the rapid development of web services, various kinds of auxiliary data (\aka side information) become available in recommender systems. Although auxiliary data is likely to contain %useful information for recommendation,
%it is difficult to model and utilize these heterogeneous and complex information in recommender systems.
Due to the flexibility in modelling data heterogeneity, heterogeneous information network (HIN) has been adopted to characterize complex and heterogeneous auxiliary data in recommender systems, called \emph{HIN based recommendation}. It is challenging to develop effective methods for HIN based recommendation in both extraction and exploitation of the information from HINs.  Most of HIN based recommendation methods rely on  path based similarity, which cannot fully mine latent structure features of users and items.
In this paper, we propose a novel heterogeneous  network embedding based approach for HIN based recommendation, called HERec.
To embed HINs, we design a meta-path based random walk strategy to generate meaningful node sequences for network embedding.
The learned node embeddings are first transformed by a set of fusion functions, and subsequently integrated into an extended matrix factorization (MF) model.
The extended MF model together with fusion functions are jointly optimized for the rating prediction task.
Extensive experiments on three real-world datasets demonstrate the effectiveness of the HERec model. Moreover, we show the capability of the HERec model for the cold-start problem, and reveal that the transformed embedding information from HINs can improve the recommendation performance.
%Extensive experiments on three real datasets show that HERec performs better than several competing baselines. %, especially in the cold-start scenario.
\end{abstract}
}

\maketitle
\IEEEdisplaynontitleabstractindextext

\IEEEraisesectionheading{\section{Introduction \label{sec-intro}}}

\IEEEPARstart{I}{n} recent years, recommender systems, which help users discover items of interest from a large resource collection, have been playing an increasingly important role in various online services~\cite{dias2008value,koren2015advances}.
Traditional recommendation methods (\eg matrix factorization) mainly aim to learn an effective prediction function for characterizing user-item interaction records (\eg user-item rating matrix). With the rapid development of web services, various kinds of auxiliary data (\aka side information) become available in recommender systems. Although auxiliary data is likely to contain useful information for recommendation~\cite{schafer2007collaborative},
it is difficult to model and utilize these heterogeneous and complex information in recommender systems.
Furthermore, it is more challenging to develop a relatively general approach to model these varying data in different systems or platforms.

\begin{figure*}[t]%[htbp]
\centering
\includegraphics[width=15cm]{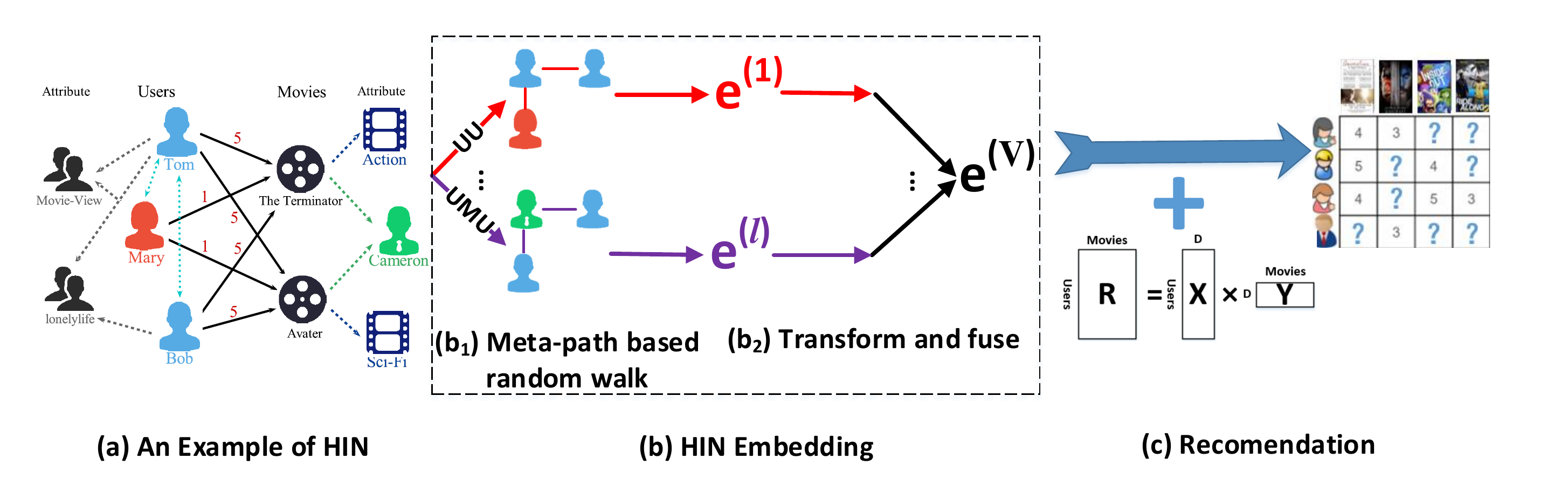}
\caption{\label{fig_framework}The schematic illustration of the proposed HERec approach.}
\end{figure*}

As a promising direction, heterogeneous information network (HIN), consisting of multiple types of nodes and links, has been proposed as a powerful information modeling method~\cite{sun2011pathsim,shi2017survey,shi2017heterogeneous}.
 Due to its flexility in modeling data heterogeneity,
 HIN has been adopted in recommender systems to characterize rich auxiliary data.
In Fig.~\ref{fig_framework}(a), we present an example for movie recommendation characterized by HINs.
We can see that the HIN contains multiple types of entities connected by different types of relations.
Under the HIN based representation, the recommendation problem can be considered as a similarity search task over the HIN~\cite{sun2011pathsim}.
Such a recommendation setting is called as \emph{HIN based recommendation}~\cite{yu2013collaborative}.
HIN based recommendation has received much attention in the literature~\cite{feng2012incorporating,yu2013collaborative,yu2014personalized,shi2015semantic,shi2016integrating,zheng2017recommendation}.
 The basic idea of most existing HIN based recommendation methods is to leverage path based semantic relatedness between users and items over HINs, \eg meta-path based similarities,
 for recommendation.

Although HIN based methods have achieved performance improvement to some extent,
there are two major problems for these methods using meta-path based similarities.
First, meta-path based similarities rely on explicit path reachability,
and may not be reliable to be used for recommendation when
path connections are sparse or noisy.
It is likely that some links in HINs are accidentally formed which do not convey meaningful semantics.
Second, meta-path based similarities mainly characterize semantic relations defined over HINs, and may not be directly applicable to
recommender systems.
It is likely that the derived path based similarities have no explicit impact on the recommendation performance in some cases.
Existing methods mainly learn a linear weighting mechanism to combine the path based similarities~\cite{shi2016integrating} or latent factors~\cite{yu2013collaborative}, which cannot learn the complicated mapping mechanism of HIN information for recommendation.
The two problems essentially reflect two fundamental issues for HIN based recommendation, namely effective information extraction and exploitation based on HINs for recommendation.
%The focus of this paper is to study how to address these two issues and improve HIN-based recommendation.

For the first issue, it is challenging to develop a way to effectively extract and represent useful information for HINs due to data heterogeneity.
Unlike previous studies using meta-path based similarities~\cite{sun2011pathsim,yu2013collaborative}, our idea is to
learn effective heterogeneous network representations for summarizing important structural characteristics and properties of HINs.
Following \cite{perozzi2014deepwalk,grover2016node2vec}, we characterize nodes from HINs with low-dimensional vectors, \ie embeddings.
Instead of relying on explicit path connection, we would like to encode useful information from HINs with latent vectors.
Compared with meta-path based similarity, the learned embeddings are in a more compact form that is easy to use and integrate.
Also, the network embedding approach itself is more resistant to sparse and noisy data.
However, most existing network embedding methods focus on homogeneous networks only consisting of a single type of nodes and links, and cannot directly deal with heterogeneous networks consisting of multiple types of nodes and links. Hence, we propose a new heterogeneous network embedding method.
Considering heterogeneous characteristics and rich semantics reflected by meta-paths, the proposed method first uses a
random walk strategy guided by meta-paths to generate node sequences. For each meta-path, we learn a unique embedding representation for a node by maximizing its co-occurrence probability with neighboring nodes in the sequences sampled according to the given meta-path. We fuse the multiple embeddings \emph{w.r.t.} different meta-paths as the output of HIN embedding.

After obtaining the embeddings from HINs, we study how to integrate and utilize such information in recommender systems.
We don't assume the learned embeddings are naturally applicable in recommender systems. Instead, we propose and explore three fusion functions to integrate multiple embeddings of a node into a single representation for recommendation, including simple linear fusion, personalized linear fusion and non-linear fusion.
These fusion functions provide flexible ways to transform HIN embeddings into useful information for recommendation.
Specially, we emphasize that personalization and non-linearity are two key points to consider for information transformation in our setting.
Finally, we extend the classic matrix factorization framework by incorporating the fused HIN embeddings.
The prediction model and the fusion function are jointly optimized for the rating prediction task.

By integrating the above two parts together, this work presents a novel HIN embedding based recommendation approach, called \emph{HERec} for short.
HERec first extracts useful HIN based information using the proposed HIN embedding method,
and then utilizes the extracted information for recommendation using the extended matrix factorization model.
We present the overall illustration for the proposed approach in Fig.~\ref{fig_framework}.
Extensive experiments on three real-world datasets demonstrate the effectiveness of the proposed approach. We also  verify the ability of HERec to alleviate cold-start problem and examine the impact of meta-paths on performance. The key contributions of this paper can be summarized as follows:

\textbullet ~We propose a heterogeneous network embedding method guided by meta-paths to uncover the semantic and structural information of heterogeneous information networks. Moreover, we propose a general embedding fusion approach to integerate different
embeddings based on different meta-paths into a single representation.

\textbullet ~We propose a novel heterogeneous information network embedding for recommendation model, called HERec for short. HERec can effectively integrate various kinds of embedding information in HIN to enhance the recommendation performance. In addition, we design a set of three flexible fusion functions to effectively transform HIN embeddings into useful information for recommendation.

\textbullet ~Extensive experiments on three real-world datasets demonstrate the effectiveness of the proposed model. Moreover, we show the capability of the proposed model for the cold-start prediction problem, and reveal that the transformed embedding information from HINs can improve the recommendation performance.

The remainder of this paper is organized as follows. Section~\ref{sec-rel} introduces the related works. Section~\ref{sec-def} describes notations used in the paper and presents some preliminary knowledge. Then, We propose the heterogeneous network embedding method and the HERec model in Section~\ref{sec-model}. Experiments and detailed analysis are reported in Section~\ref{sec-exp}. Finally, we conclude the paper in Section~\ref{sec-con}.

\begin{figure*}[t]
\centering
\subfigure[Douban Movie]{
\begin{minipage}[b]{0.3\textwidth}
\includegraphics[width=1\textwidth]{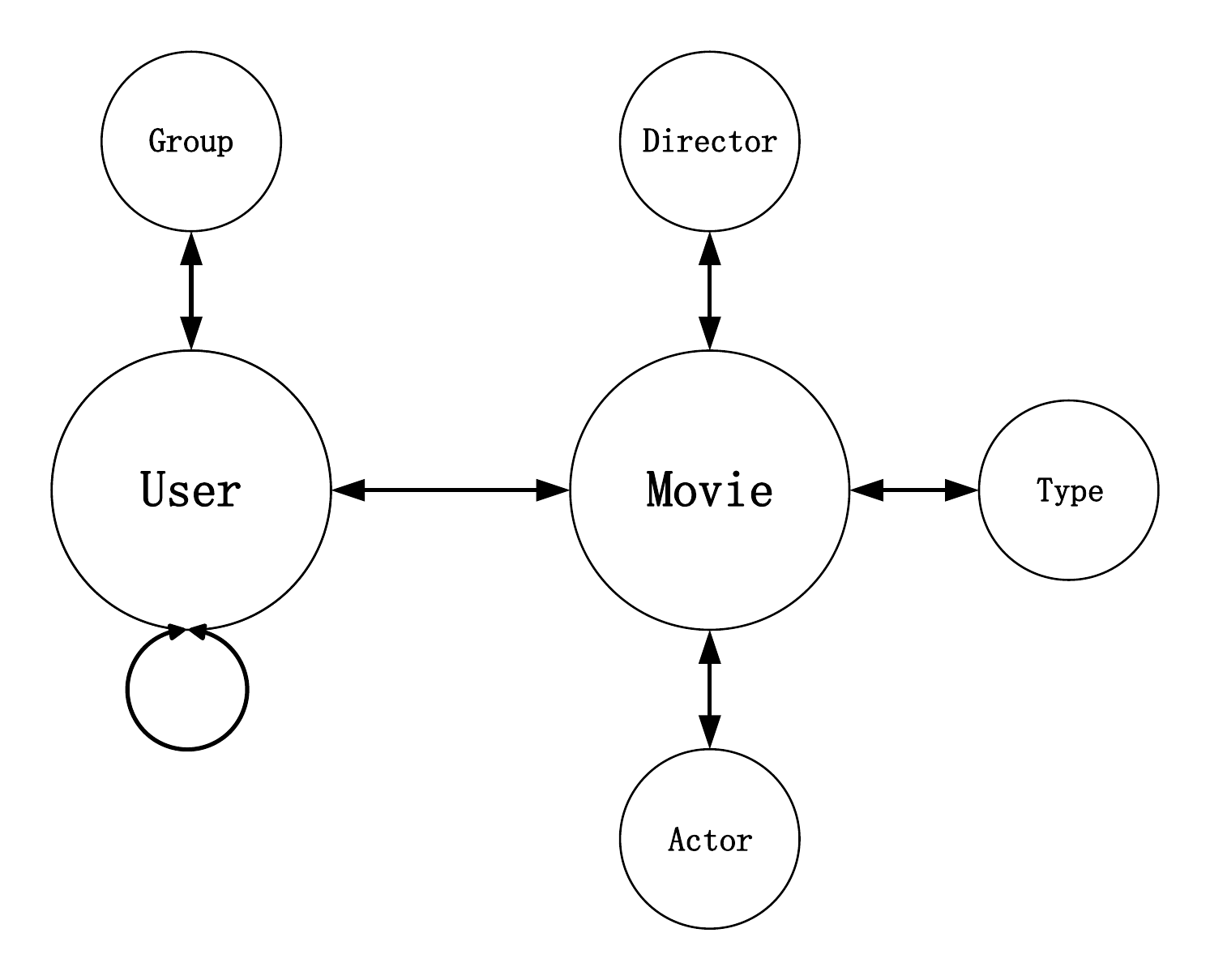}
\end{minipage}
}
\subfigure[Douban Book]{
\begin{minipage}[b]{0.3\textwidth}
\includegraphics[width=1\textwidth]{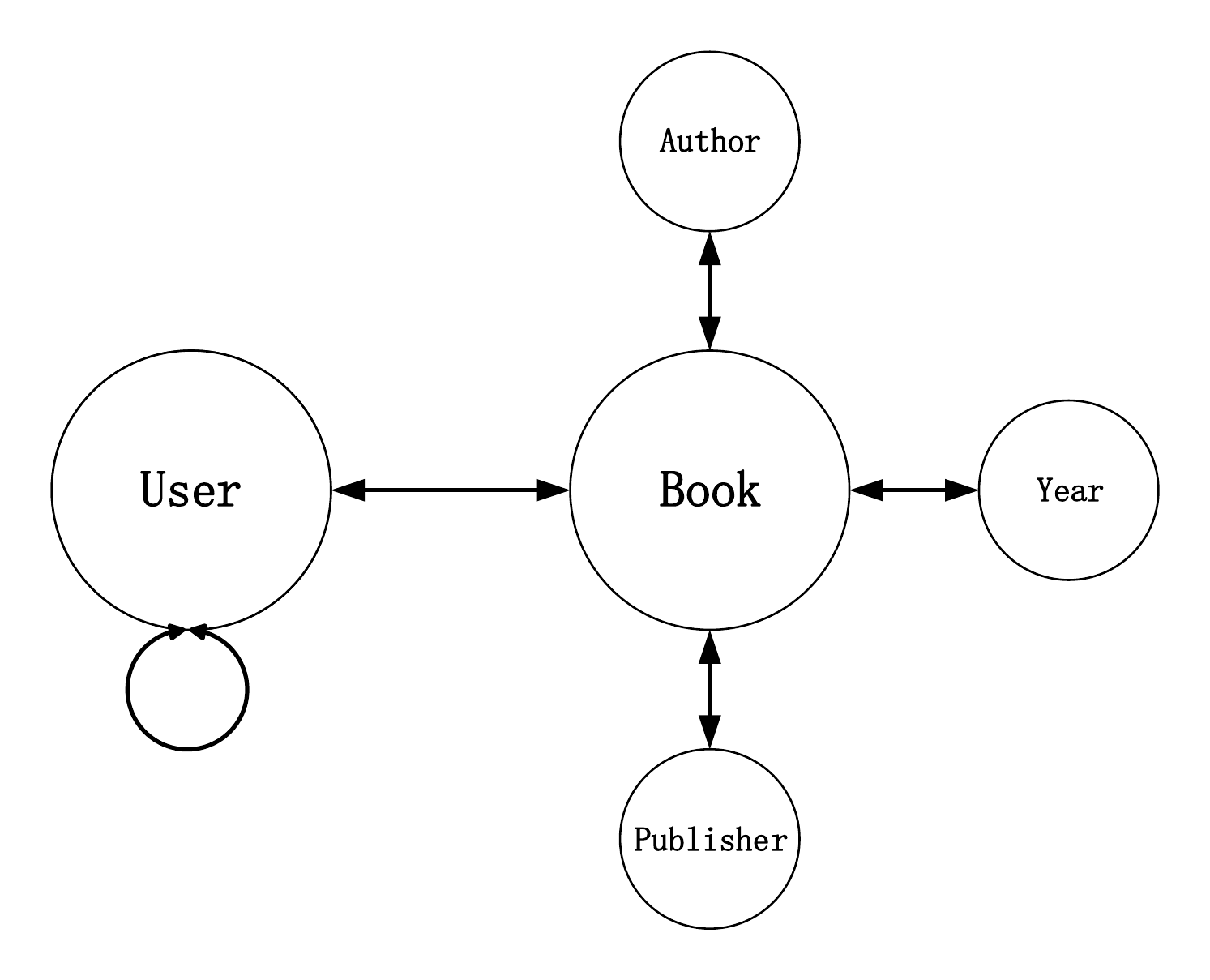}
\end{minipage}
}
\subfigure[Yelp]{
\begin{minipage}[b]{0.3\textwidth}
\includegraphics[width=1\textwidth]{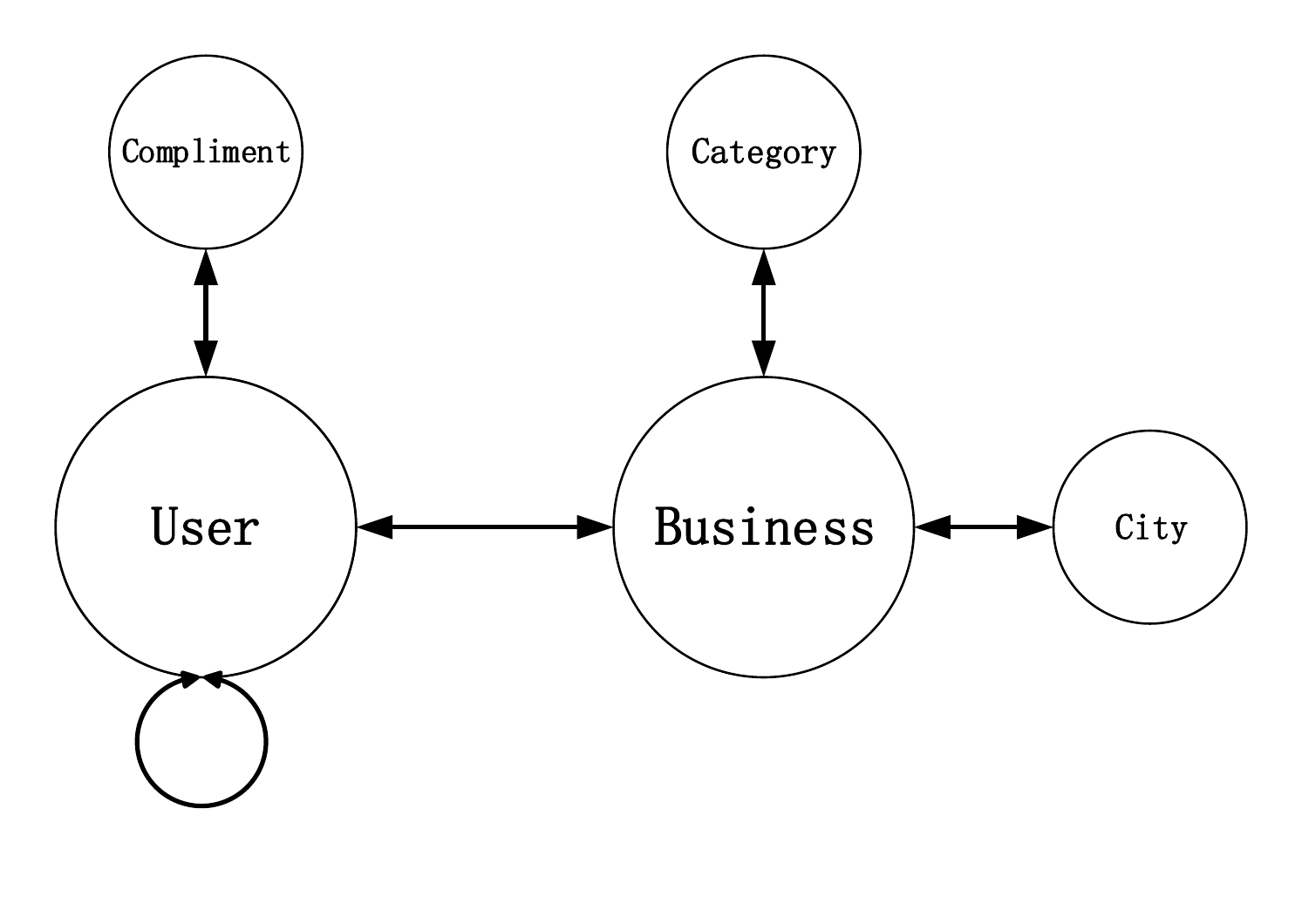}
\end{minipage}
}
\caption{\label{fig_schema}Network schemas of heterogeneous information networks for the used three datasets.
In our task, users and items are our major focus, denoted by large-sized circles, while the other attributes are denoted by small-sized circles. }

\end{figure*}

\section{Related Work \label{sec-rel}}
In this section, we will review the related studies in three aspects, namely recommender systems, heterogeneous information networks and network embedding.

In the literature of recommender systems, early works mainly adopt collaborative filtering (CF) methods to utilize historical interactions for recommendation~\cite{schafer2007collaborative}. Particularly, the matrix factorization approach~\cite{koren2009matrix,shi2012adaptive} has shown its effectiveness and efficiency in many applications, which factorizes user-item rating matrix into two low rank user-specific and item-specific matrices, and then utilizes the factorized matrices to make further predictions~\cite{koren2015advances}. Since CF methods usually suffer from cold-start problem, many works~\cite{yin2013lcars,feng2012incorporating,hong2013co} attempt to leverage additional information to improve recommendation performance. For example, Ma et al.~\cite{ma2011recommender} integrate social relations into matrix factorization in recommendation. Ling et al.~\cite{ling2014ratings} consider the information of both ratings and reviews and propose a unified model to combine content based filtering with collaborative filtering for rating prediction task. Ye et al.~\cite{ye2011exploiting} incorporate user preference, social influence and geographical influence in the recommendation and propose a unified POI recommendation framework. More recently, Sedhain et al.~\cite{sedhain2017low} explain drawbacks of three popular cold-start models~\cite{gantner2010learning,krohn2012multi,sedhain2014social} and further propose a learning based approach for the cold-start problem to leverage social data via randomised SVD. And many works begin to utilize deep models ($\eg$ convolutional neural network, auto encoder) to exploit text information~\cite{zheng2017joint}, image information~\cite{he2016vbpr} and network structure information~\cite{zhang2016collaborative} for better recommendation. In addition, there are also some typical frameworks focusing on incorporating auxiliary information for recommendation. Chen et al.~\cite{chen2012svdfeature} propose a typical SVDFeature framework to efficiently solve the feature based matrix factorization. And Rendle~\cite{rendle2010factorization} proposes factorization machine, which is a generic approach to combine the generality of feature engineering.

%In the literature of recommender systems, early works mainly adopt collaborative filtering (CF) methods (\eg matrix factorization) to utilize historical interactions for recommendation \cite{schafer2007collaborative,koren2015advances}.
%Since CF methods usually suffers from cold-start problem, many works attempted to leverage additional information to improve recommendation performance, such as social information \cite{he2010social}, location information \cite{yin2013lcars}, and heterogeneous information \cite{feng2012incorporating}.

As a newly emerging direction, heterogeneous information network~\cite{shi2017survey} can naturally model complex objects and their rich relations in recommender systems, in which objects are of different types and links among objects represent different relations~\cite{sun2013mining}. And several path based similarity measures~\cite{lao2010relational,sun2011pathsim,shi2014hetesim} are proposed to evaluate the similarity of objects in heterogeneous information network. Therefore, some researchers have began to be aware of the importance of HIN based recommendation. Wang et al.~\cite{feng2012incorporating} propose the OptRank method to alleviate the cold-start problem by utilizing heterogeneous information contained in social tagging system. Furthermore, the concept of meta-path is introduced into hybrid recommender systems~\cite{yu2013recommendation}. Yu et al.~\cite{yu2013collaborative} utilize meta-path based similarities as regularization terms in the matrix factorization framework. Yu et al.~\cite{yu2014personalized} take advantage of different types of entity relationships in heterogeneous information network and propose a personalized recommendation framework for implicit feedback dataset. Luo et al.~\cite{luo2014hete} propose a collaborative filtering based social recommendation method using heterogeneous relations. More recently, Shi et al.~\cite{shi2015semantic} propose the concept of weighted heterogeneous information network and design a meta-path based collaborative filtering model to flexibly integrate heterogeneous information for personalized recommendation. In \cite{shi2016integrating,zheng2016dual,zheng2017recommendation}, the similarities of users and items are both evaluated by path based similarity measures under different semantic meta-paths and a matrix factorization based on dual regularization framework is proposed for rating prediction. Most of HIN based methods rely on the path based similarity, which may not fully mine latent features of users and items on HINs for recommendation.

On the other hand, network embedding has shown its potential in structure feature extraction and has been successfully applied in many data mining tasks~\cite{hoff2002latent,yan2007graph}, such as classification~\cite{tu2016max}, clustering~\cite{wei2017cross,cao2016deep} and recommendation~\cite{liang2016factorization}. Deepwalk~\cite{perozzi2014deepwalk} combines random walk and skip-gram to learn network representations. Furthermore, Grover and Leskovec~\cite{grover2016node2vec} propose a more flexible network embedding framework based on a biased random walk procedure. In addition, LINE \cite{tang2015line} and SDNE~\cite{wang2016structural} characterize the second-order link proximity, as well as neighbor relations. Cao et al.~\cite{cao2015grarep} propose the GraRep model to capture higher-order graph proximity for network representations. Besides leaning network embedding from only the topology,  there are also many works~\cite{pan2016tri,yang2015network,zhang2016homophily} leveraging node content information and other available graph information for the robust representations. Unfortunately, most of network embedding methods focus on homogeneous networks, and thus they cannot directly be applied for heterogeneous networks. Recently, several works~\cite{chang2015heterogeneous,tang2015pte,xu2017embedding,chen2017task,dong2017metapath2vec} attempt to analyze heterogeneous networks via embedding methods. Particularly, Chang et al.~\cite{chang2015heterogeneous} design a deep embedding model to capture the complex interaction between the heterogeneous data in the network. Xu et al.~\cite{xu2017embedding} propose a EOE method to encode the intra-network and inter-network edges for the coupled heterogeneous network. Dong et al.~\cite{dong2017metapath2vec} define the neighbor of nodes via meta-path and learn the heterogeneous embedding by skip-gram with negative sampling. Although these methods can learn network embeddings in various heterogeneous network, their representations of nodes and relations may not be optimum for recommendation.

%Pan et al.\cite{pan2016tri} propose the TriDNR model to learn optimal node representation with node structure information, node content and node labels. Yang et al.~\cite{yang2015network} propose the TADW model to incorporate text features of vertices into network representation based on matrix factorization framework. Zhang et al.~\cite{zhang2016homophily} consider neighbors homophily, topology structure and node content during network representation learning based on matrix decomposition. Most of network embedding methods focus on homogeneous networks, and thus they cannot directly be applied for heterogeneous networks. Although several works~\cite{chang2015heterogeneous,tang2015pte,xu2017embedding,chen2017task,dong2017metapath2vec} attempt to analyze heterogeneous networks via embedding methods, their representations of nodes and relations may not be suitable for recommendation.

To our knowledge, it is the first attempt which adopts the network embedding approach to extract useful information from heterogeneous information network and leverage such information for rating prediction. The proposed approach utilizes the flexibility of HIN for modeling complex heterogeneous context information, and meanwhile borrows the capability of network embedding for learning effective information representation.
The final rating prediction component further incorporates a transformation mechanism implemented by three flexible functions to utilize the learned information from network embedding.

\section{Preliminary \label{sec-def}}

A heterogeneous information network is a special kind of information network, which either contains multiple types of objects or multiple types of links.
\begin{myDef}
\textbf{Heterogeneous information network}~\cite{sun2012mining}. A HIN is denoted as $\mathcal{G} = \{\mathcal{V}, \mathcal{E}\}$ consisting of a object set $\mathcal{V}$ and a link set $\mathcal{E}$. A HIN is also associated with an object type mapping function $\phi: \mathcal{V} \rightarrow \mathcal{A}$ and a link type mapping function $\psi: \mathcal{E} \rightarrow \mathcal{R}$. $\mathcal{A}$ and $\mathcal{R}$ denote the sets of predefined object and link types, where $|\mathcal{A}| + |\mathcal{R}| > 2$.
\end{myDef}

The complexity of heterogeneous information network drives us to provide the meta level ($\eg$ schema-level) description for understanding the object types and link types better in the network. Hence, the concept of network schema is proposed to describe the meta structure of a network.
\begin{myDef}
\textbf{Network schema}~\cite{sun2013mining,sun2009ranking}. The network schema is denoted as $\mathcal{S} = (\mathcal{A}, \mathcal{R})$. It is a meta template
for an information network $\mathcal{G} = \{\mathcal{V}, \mathcal{E}\}$ with the object type mapping $\phi: \mathcal{V} \rightarrow \mathcal{A}$ and the link type mapping $\psi: \mathcal{E} \rightarrow \mathcal{R}$, which is a directed graph defined over object types $\mathcal{A}$, with edges as relations from $\mathcal{R}$.
\end{myDef}

\begin{exmp}
As shown in Fig.~\ref{fig_framework}(a), we have represented the setting of movie recommender systems by HINs.
We further present its corresponding network schema in Fig.~\ref{fig_schema}(a), consisting of multiple types of
objects, including User ($U$), Movie ($M$), Director ($D$). There exist different types of links between objects to represent different
relations. A user-user link indicates the friendship between two users, while a user-movie link indicates the rating relation.
Similarly, we present the schematic network schemas for book and business recommender systems in Fig.~\ref{fig_schema}(b) and Fig.~\ref{fig_schema}(c) respectively.
\end{exmp}

In HINs, two objects can be connected via different semantic paths, which are called meta-paths.

\begin{myDef}
\textbf{Meta-path}~\cite{sun2011pathsim}. A meta-path $\rho$ is defined on a network schema $\mathcal{S} = (\mathcal{A}, \mathcal{R})$  and is denoted as a path in the form of $A_1 \xrightarrow{R_1} A_2 \xrightarrow{R_2} \cdots \xrightarrow{R_l} A_{l+1}$ (abbreviated as $A_1A_2 \cdots A_{l+1}$), which describes a composite relation $R = R_1 \circ R_2 \circ \cdots \circ R_l$ between object $A_1$ and $A_{l+1}$, where $\circ$ denotes the composition operator on relations.
\end{myDef}

\begin{exmp}
Taking Fig.~\ref{fig_schema}(a) as an example, two objects can be connected via multiple meta-paths, $\eg$ ``User - User" ($UU$) and  ``User - Movie - User" ($UMU$). Different meta-paths usually convey different semantics. For example, the $UU$ path indicates friendship between two users, while the $UMU$ path indicates the co-watch relation between two users, \ie they have watched the same movies. As will be seen later, the detailed meta-paths used in this work is summarized in Table~\ref{tab_metapath}.
\end{exmp}
%Taking Fig. 1(a) as an example, we construct a HIN to model the movie recommendation setting,
% which consists of multiple types of objects (\eg User ($U$), Movie ($M$), Director ($D$)) and links (\eg social relation between users and rating relation between users and movies).
%In this example, two objects can be connected via multiple meta-paths, \eg
% ``User-User" ($UU$) and  ``User-Movie-User" ($UMU$).
% Different meta-paths often convey different semantics. For example, the $UU$ path indicates friendship between two users, while
% the $UMU$ path indicates the two users have watched the same movies. As a major technical approach, meta-path-based search and mining methods have been extensively studied in HINs %\cite{shi2017heterogeneous}.

Recently, HIN has become a mainstream approach to model various complex interaction systems \cite{shi2017survey}. Specially, it has been adopted in recommender systems for characterizing complex and heterogenous recommendation settings.

\begin{myDef}
\textbf{HIN based recommendation}. In a recommender system, various kinds of information can be modeled by a HIN $\mathcal{G} = \{\mathcal{V}, \mathcal{E}\}$. On recommendation-oriented HINs, two kinds of entities (\ie users and items) together with the relations between them (\ie rating relation) are our focus.
Let $\mathcal{U}\subset \mathcal{V}$
and $\mathcal{I}\subset \mathcal{V}$ denote the sets of users and items respectively, a triplet $\langle u, i, r_{u,i}\rangle$  denotes a record that a user $u$ assigns a rating of $r_{u,i}$ to  an item $i$, and $\mathcal{R}=\{\langle u, i, r_{u,i}\rangle\}$ denotes the set of rating records.
We have $\mathcal{U}\subset \mathcal{V}$, $\mathcal{I}\subset \mathcal{V}$ and $\mathcal{R}\subset \mathcal{E}$.
Given the HIN $\mathcal{G} = \{\mathcal{V}, \mathcal{E}\}$, the goal is to predict the rating score $r_{u,i'}$ of $u\in \mathcal{U}$ to a non-rated item $i'\in \mathcal{I}$.
\end{myDef}

Several efforts have been made for HIN based recommendation. Most of these works mainly leverage meta-path based similarities to enhance
the recommendation performance~\cite{yu2013collaborative,yu2014personalized,shi2015semantic,shi2016integrating}. Next, we will present a new heterogeneous network embedding based approach to this task, which is able to effectively
exploit the information reflected in HINs. The notations we will use throughout the article are summarized in Table~\ref{tabl_notations}.

\begin{table}[htbp]
\centering
\caption{Notations and explanations.}\label{tabl_notations}{
\begin{tabular}{c||c}
\hline
{Notation} & {Explanation}\\
\hline
\hline
{$\mathcal{G}$}&{heterogeneous information network}\\
\hline
{$\mathcal{V}$} & {object set} \\
\hline
{$\mathcal{E}$} & {link set} \\
\hline
{$\mathcal{S}$} & {network schema} \\
\hline
{$\mathcal{A}$} & {object type set } \\
\hline
{$\mathcal{R}$} & {link type set} \\
\hline
{$\mathcal{U}$} & {user set} \\
\hline
{$\mathcal{I}$} & {item set} \\
\hline
{$\widehat{r_{u,i}}$} & {predicted rating user $u$ gives to item $i$}\\
\hline
{$\bm{e}_v$} & {low-dimensional representation of node $v$} \\
\hline
{$\mathcal{N}_u$} & {neighborhood of node $u$} \\
\hline
{$\rho$} & {a meta-path} \\
\hline
{$\mathcal{P}$} & {meta-path set} \\
\hline
{$\bm{e}^{(U)}_u, \bm{e}^{(I)}_i$} & {final representations of  user $u$, item $i$}\\
\hline
{$d$} & {dimension of HIN embeddings} \\
\hline
{$D$} & {dimension of latent factors} \\
\hline
{$\mathbf{x}_u, \mathbf{y}_i$} & {latent factors of user $u$, item $i$} \\
\hline
{$\bm{{\gamma}}^{(U)}_u$, $\bm{{\gamma}}_i^{(I)}$} & {latent factors for pairing HIN embedding of user $u$, item $i$} \\
\hline
{$\alpha$, $\beta$} & {parameters for integrating HIN embeddings} \\
\hline
{$\mathbf{M}^{(l)}$} & { transformation matrix w.r.t the $l$-th mete-path} \\
\hline
{$\bm{b}^{(l)}$} & { bias vector w.r.t the $l$-th mete-path} \\
\hline
{$w^{(l)}_u$} & {preference weight of user $u$ over the $l$-th meta-path} \\
\hline
{$\bm{\Theta}^{(U)}$, $\bm{\Theta}^{(I)}$} & {parameters of fusion functions for users, items} \\
\hline
{$\lambda$} & {regularization parameter} \\
\hline
{$\eta$} & {learning rate} \\
\hline
\end{tabular}}
\end{table}

\section{The Proposed Approach \label{sec-model}}

In this section, we present a \emph{H}eterogeneous network \emph{E}mbedding based  approach for \emph{Rec}emendation, called  \emph{HERec}.
For addressing the two issues introduced in Section 1, the proposed HERec approach consists of two major components.
First, we propose a new heterogeneous network embedding method to learn the user/item embeddings from HINs.
Then, we extend the classic matrix factorization framework by incorporating the learned embeddings using a flexible set of fusion functions.
We present an overall schematic illustration of the proposed approach in Fig.~\ref{fig_framework}.
After the construction of the HINs~(Fig.~\ref{fig_framework}(a)), two major steps are presented,  namely HIN embedding~(Fig.~\ref{fig_framework}(b)) and recommendation~(Fig.~\ref{fig_framework}(c)). Next, we will present detailed illustration of the proposed approach.

\subsection{Heterogeneous Network Embedding}
Inspired by the recent progress on network embedding~\cite{perozzi2014deepwalk,grover2016node2vec}, we adopt the representation learning method to extract and represent useful information of HINs for recommendation. Given a HIN $\mathcal{G} = \{\mathcal{V}, \mathcal{E}\}$, our goal is to learn a low-dimensional representation $\bm{e}_v \in \mathbb{R}^d$ (\aka embedding) for each node $v \in \mathcal{V}$.
The learned embeddings are expected to highly summarize informative characteristics, which are likely to be useful in recommender systems represented on HINs.
Compared with meta-path based similarity~\cite{sun2011pathsim,shi2015semantic}, it is much easier to use and integrate the learned representations in subsequent procedures.

However, most of the existing network embedding methods mainly focus on homogeneous networks, which are not able to effectively model heterogeneous networks. For instance, the pioneering study \emph{deepwalk}~\cite{perozzi2014deepwalk} uses random walk to generate node sequences, which
cannot discriminate nodes and edges with different types. Hence, it requires a more principled way to traverse the HINs and generate meaningful node sequences.

%In this section, we will present the details of heterogeneous network embedding. The basic idea is to employ a meta-path based random walk to generate a sequence of nodes, and then we maximize the co-occurrence probability of nodes to obtain the latent feature representation of nodes for one single meta-path, which reflect the structural characteristics of nodes from one perspective. Furthermore, we propose a flexible embedding fusion framework to combine the latent feature representation of nodes from different meta-paths.

\subsubsection{Meta-path based Random Walk}
To generate meaningful node sequences, the key is to design an effective walking strategy that is able to capture the complex semantics reflected in HINs. In the literature of HINs, meta-path is an important concept to characterize the semantic patterns for HINs~\cite{sun2011pathsim}.
Hence, we propose to use the meta-path based random walk method to generate node sequences. Giving a heterogeneous network $\mathcal{G} = \{\mathcal{V}, \mathcal{E}\}$ and a meta-path $\rho : A_1 \xrightarrow{R_1} \cdots A_t \xrightarrow{R_t} A_{t+1} \cdots \xrightarrow{R_l} A_{l+1}$, the walk path is generated according to the following distribution:

\begin{eqnarray}\label{eq-rw}
&&P({n_{t+1}=x |n_{t}=v, \rho})\\
&=&\begin{cases}
\frac{1}{|\mathcal{N}^{A_{t+1}}(v)|}, &\text{($v, x$) $\in$ $\mathcal{E}$ and $\phi(x) = A_{t+1}$};\nonumber\\
0,& \text{otherwise},
\end{cases}
\end{eqnarray}
where $n_t$ is the $t$-th node in the walk, $v$ has the type of $A_t$, and $\mathcal{N}^{(A_{t+1})}(v)$ is the first-order neighbor set for node $v$ with the type of $A_{t+1}$. A walk will follow the pattern of a meta-path repetitively until it reaches the pre-defined length.

\begin{exmp}
We still take Fig.~\ref{fig_framework}(a) as an example, which represents the heterogeneous information network of movie recommender systems.
Given a meta-path $UMU$, we can generate two sample walks (\ie node sequences) by starting with the user node of \emph{Tom}: (1) Tom$_{User}$ $\rightarrow$ The Terminator$_{Movie}$ $\rightarrow$  Mary$_{User}$, and (2) Tom$_{User}$ $\rightarrow$  Avater$_{Movie}$ $\rightarrow$  Bob$_{User}$ $\rightarrow$  The Terminator$_{Movie}$ $\rightarrow$  Mary$_{User}$. Similarly, given the meta-path $UMDMU$, we can also generate another node sequence: Tom$_{User}$ $\rightarrow$  The Terminator$_{Movie}$ $\rightarrow$  Cameron$_{Director}$  $\rightarrow$  Avater$_{Movie}$  $\rightarrow$ Mary$_{User}$. It is intuitive to see that these meta-paths can lead to meaningful node sequences corresponding to different semantic relations.
\end{exmp}

\begin{figure}[t]%[htbp]
\centering
\includegraphics[width=9cm]{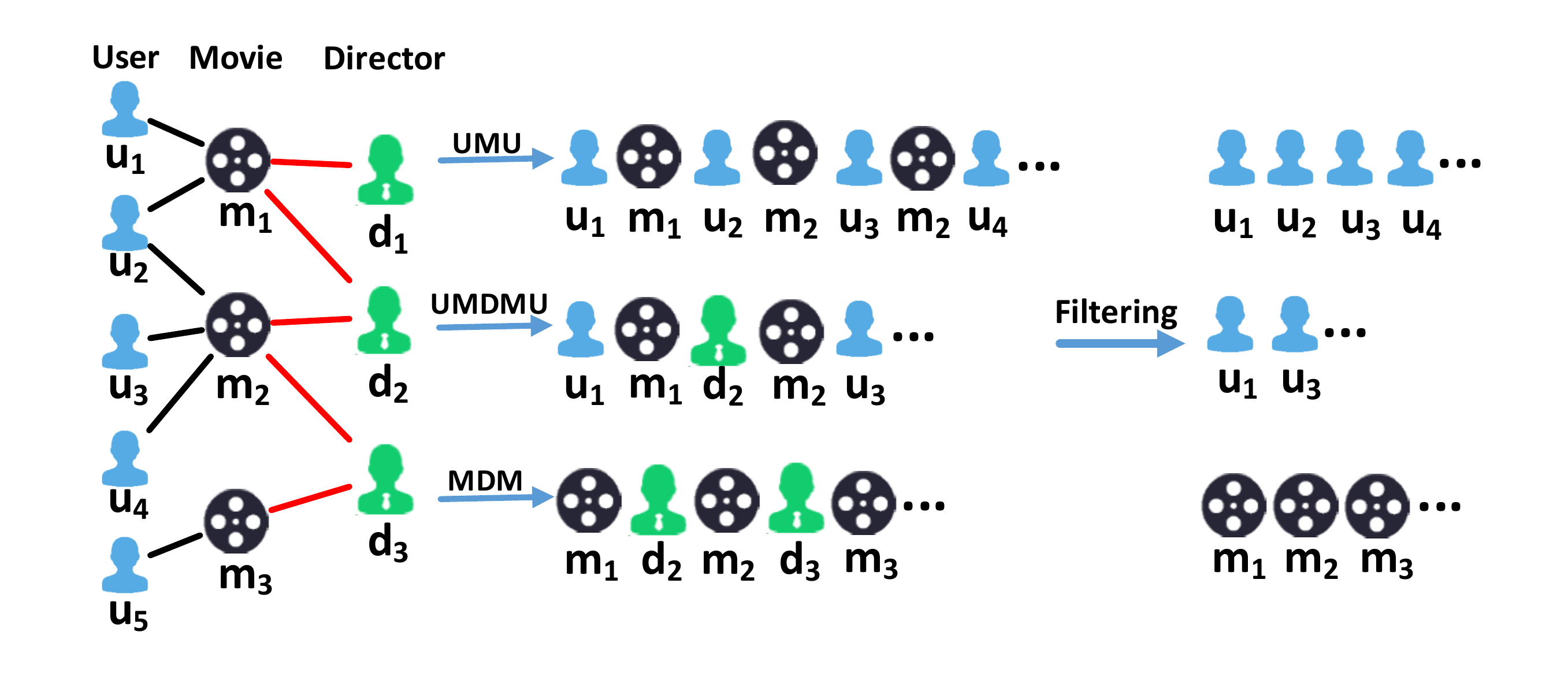}
\caption{\label{fig_randwalk}An illustrative example of the proposed meta-path based random walk. We first perform random walks guided by some selected meta-paths, and then filter the node sequences \emph{not with} the user type or item type.}
\end{figure}

\subsubsection{Type Constraint and Filtering}
Since our goal is to improve the recommendation performance, the main focus is to learn effective representations for users and items,
while objects with other types are of less interest in our task.
Hence, we only select meta-paths starting with \emph{user type} or \emph{item type}.
Once a node sequence has been generated using the above method, it is likely to contain nodes with different types.
We further remove the nodes with a type different from the starting type.
In this way, the final sequence will only consist of nodes with the starting type.
Applying type filtering to node sequences has two benefits. First, although node sequences are constructed using meta-paths with heterogeneous types, the final representations
are learned using homogeneous neighborhood. We embed the nodes with the same type in the same space, which relaxes the challenging goal of representing all the heterogeneous objects in a unified space.
Second, given a fixed-length window, a node is able to utilize more homogeneous neighbors that are more likely to be relevant than others with different types.
%We present an illustrative example of the proposed meta-path random walk method in Fig. \ref{fig_randwalk}.

\begin{exmp}
As shown in Fig.~\ref{fig_randwalk}, in order to learn effective representations for users and items, we only consider the meta-paths in which the starting type is \emph{user type} or \emph{item type}. In this way, we can derive some meta-paths, such as $UMU$, $UMDMU$ and $MUM$.
 Take the meta-path of $UMU$ as an instance. We can generate a sampled sequence  ``$u_1 \rightarrow m_1 \rightarrow u_2 \rightarrow m_2 \rightarrow u_3 \rightarrow m_2 \rightarrow u_4$" according to Eq.~\ref{eq-rw}. Once a sequence has been constructed, we further remove the nodes with a different type compared with the starting node. In this way, we finally obtain a homogeneous node sequence ``$u_1 \rightarrow u_2 \rightarrow u_3 \rightarrow u_4$".
\end{exmp}

 The connections between homogeneous nodes are essentially constructed via the heterogeneous neighborhood nodes. After this step, our next focus will be how to learn effective representations for homogeneous sequences.

\subsubsection{Optimization Objective}
Given a meta-path, we can construct the neighborhood $\mathcal{N}_u$ for node $u$ based on co-occurrence in a fixed-length window.
Following node2vec~\cite{grover2016node2vec}, we can learn the representations of nodes to optimize the following objective:
\begin{equation}
\max_f \sum_{u \in \mathcal{V}}{\log Pr(\mathcal{N}_u | f(u))},
\end{equation}
where $f: \mathcal{V} \rightarrow \mathbb{R}^d$ is a function (aiming to learn) mapping each node to $d$-dimensional feature space, and $\mathcal{N}_u  \subset \mathcal{V}$ represents the neighborhood of node $u$, \emph{w.r.t.} a specific meta-path. We can learn the embedding mapping function $f(\cdot)$ by applying stochastic gradient descent (SGD) to optimize this objective. %The embedding algorithm is shown in Algorithm 1.
A major difference between previous methods and ours lies in the construction of $\mathcal{N}_u$. Our method selects homogeneous neighbors using meta-path based random walks. The whole algorithm framework is shown in Algorithm~\ref{alg_embedding}. %where $path\_constraint\_filter($path$)$ is the procedure performing the type constraint and filtering as discussed in the previous subsection.

\begin{algorithm}[htb]
\caption{HIN embedding algorithm for a single meta-path.}
\label{alg_embedding}
\begin{algorithmic}[1]
\Require
the heterogeneous information network $\mathcal{G} = \{\mathcal{V}, \mathcal{E}\}$;
the given meta-path $\rho$;
the target node type $A_t$;
the dimension of embedding $d$;
the walk length $wl$;
the neighborhood size $ns$;
the number of walks per node $r$.

\Ensure
The embedding of target node type w.r.t the single meta-path, denoted by $\bm{e}$
\State Initialize $e$ by standard normal distribution;
\State $paths = []$;
\For {each $v$ $\in$ $\mathcal{V}$ and $\phi(v) == A_t$}
    \For {$i$ = 1 to $r$}
        \State $path = []$;
        \While {$wl > 0$}
            \State walk to node $x$ according to Eq.~\ref{eq-rw};
            \If {$\phi(x) == A_t$}
                \State append node $x$ into $path$;
                \State $wl \leftarrow wl - 1$;
            \EndIf
         \EndWhile
         %\State path\_constraint\_filter($path$);
         \State Add $path$ to $paths$;
     \EndFor
\EndFor

\State $\bm{e} = SGD(paths, d, ns)$;
\State \Return {$\bm{e}$}.
\end{algorithmic}
\end{algorithm}

\subsubsection{Embedding Fusion}
Heterogeneous network embedding provides a general way to extract useful information from HINs.
For our model, given a node $v \in \mathcal{V}$, we can obtain a set of representations $\{ \bm{e}^{(l)}_v \}_{l=1}^{|\mathcal{P}|}$,
where $\mathcal{P}$ denotes the set of meta-paths, and $\bm{e}^{(l)}_v$ denotes the representation of $v$ \emph{w.r.t.} the $l$-th meta-path.
It requires a principled fusion way to transform node embeddings into a more suitable form that is useful to improve recommendation performance.
Existing studies usually adopt a linear weighting mechanism to combine the information mined from HINs (\eg meta-path based similarities), which may not be capable of deriving effective information representations for recommendation.
Hence, we propose to use a general function $g(\cdot)$, which aims to fuse the learned node embeddings for users and items:

\begin{eqnarray}\label{eq-eui}
\bm{e}^{(U)}_u &\leftarrow& g(\{\bm{e}^{(l)}_u\}),\\
\bm{e}^{(I)}_i &\leftarrow& g(\{\bm{e}^{(l)}_i\}),\nonumber
\end{eqnarray}
where $\bm{e}^{(U)}_u$ and $\bm{e}^{(I)}_i$ are the final representations for a user $u$ and an item $i$ respectively, called \emph{HIN embeddings}.
Since users and items are our focus, we only learn the embeddings for users and items.
At the current stage, we do not specify the form of function $g(\cdot)$. Instead, we believe that
a good fusion function should be learned according to the specific task. Hence, we leave the formulation and optimization of the fusion function
in our recommendation model.

\subsection{Integrating Matrix Factorization with Fused HIN Embedding for Recommendation}
Previously, we have studied how to extract and represent useful information from HINs for recommendation.
With HIN embedding, we can obtain user embeddings $\{\bm{e}^{(U)}_u\}_{u \in \mathcal{U}}$ and item embeddings $\{\bm{e}^{(I)}_i\}_{i \in \mathcal{I}}$, which are further specified by a function $g(\cdot)$ that is to learn. Now we study how to utilize the learned embeddings for recommendation.

\subsubsection{Rating Predictor}
We build our rating predictor based on the classic matrix factorization (MF) model~\cite{mnih2008probabilistic}, which factorizes the user-item rating matrix into user-specific and item-specific matrices. In MF, the rating of a user $u$ on an item $i$ is simply defined as follows:

\begin{equation}\label{eq-mf}
\widehat{r_{u,i}} = \mathbf{x}_u^{\top}\cdot \mathbf{y}_i,
\end{equation}
where $\mathbf{x}_u \in \mathbb{R}^{D}$ and $\mathbf{y}_i \in \mathbb{R}^{D}$ denote the latent factors corresponding to user $u$ and item $i$.
Since we have also obtained the representations for user $u$ and item $i$, we further incorporate them into the rating predictor as below

\begin{equation}\label{eq-predictor}
\widehat{r_{u,i}} = \mathbf{x}_u^{\top}\cdot \mathbf{y}_i + \alpha \cdot {\bm{e}_u^{(U)}}^{\top}\cdot\bm{{\gamma}}_i^{(I)}  + \beta \cdot {\bm{{\gamma}}^{(U)}_u}^{\top}\cdot {\bm{e}_i^{(I)}},
\end{equation}
where $\bm{e}_u^{(U)}$ and $\bm{e}_i^{(I)}$ are the fused embeddings,
$\bm{{\gamma}}^{(U)}_u$ and $\bm{{\gamma}}_i^{(I)}$ are user-specific and item-specific latent factors to pair with the HIN embeddings
$\bm{e}_u^{(U)}$ and $\bm{e}_i^{(I)}$ respectively, and $\alpha$ and $\beta$ are the tuning parameters to integrate the three terms.
For Eq.~\ref{eq-predictor}, we need to note two points. First, $\bm{e}_u^{(U)}$ and $\bm{e}_i^{(I)}$ are the output of function $g(\cdot)$ in Eq.~\ref{eq-eui}.
We assume that the derived embeddings after transformation by function $g(\cdot)$ are applicable
in MF. Second, we don't directly pair $\bm{e}_u^{(U)}$ with $\bm{e}_i^{(I)}$.
Recall the proposed embedding method indeed characterizes the relatedness between objects with the same type.
We incorporate new latent factors $\bm{{\gamma}}^{(U)}_u$ and $\bm{{\gamma}}_i^{(I)}$ to relax
the assumption that $\bm{e}_u^{(U)}$ and $\bm{e}_i^{(I)}$ have to be in the same space, which increases the flexility to the prediction model.

\subsubsection{Setting the Fusion Function}
Previously, we assume the function $g(\cdot)$ has been given in a general form.
Now, we study how to set the fusion function, which transforms HIN embeddings into a form that is useful in recommender systems.
We only discuss the function for fusing user embeddings, and it is similar to fuse item embeddings.
We propose to use three fusion functions to integrate embeddings:

\begin{itemize}
\item Simple linear fusion. We assume each user has the same preference for each meta-path, and therefore, we assign each meta-path with a unified weight ($\ie$ average value) for each user. Moreover, we linearly transform embeddings to target space.
\begin{equation}\label{eq-slf}
g(\{\bm{e}^{(l)}_u\}) = \frac{1}{|\mathcal{P}|}\sum_{l=1}^{|\mathcal{P}|}{(\mathbf{M}^{(l)} \bm{e}^{(l)}_u+\bm{b}^{(l)})},
\end{equation}
where $\mathcal{P}$ is the set of considered meta-paths, $\mathbf{M}^{(l)} \in \mathbb{R}^{D\times d}$ and $\bm{b}^{(l)}  \in \mathbb{R}^{D}$ are the transformation matrix and bias vector \emph{w.r.t.} the $l$-th meta-path.
\item Personalized linear fusion. The simple linear fusion cannot model users' personalized preference over the meta-paths. So we further assign each user with a weight vector on meta-paths, representing user's personalized preference for each meta-path. It is more reasonable that each user has his/her personal interest preference in many real applications.
\begin{equation}\label{eq-plf}
g(\{\bm{e}^{(l)}_u\}) = \sum_{l=1}^{|\mathcal{P}|}{w^{(l)}_u (\mathbf{M}^{(l)} \bm{e}^{(l)}_u+\bm{b}^{(l)})},
\end{equation}
where $w^{(l)}_u$ is the preference weight of user $u$ over the $l$-th meta-path.
\item Personalized non-linear fusion. Linear fusion has limited expressive power in modeling complex data relations.
Hence, we use non-linear function to enhance the fusion ability.
\begin{equation}\label{eq-pnlf}
%F(\mathbf{X}^{(u)}_i) = \sigma(\sum_{l=1}^{|\mathcal{P}^{(u)}|}{p^{(u)}_{il} \sigma (\mathbf{W}^{(u)}_l\mathbf{X}_{il}+\mathbf{b}^{(u)}_l)}).
g(\{\bm{e}^{(l)}_u\}) = \sigma\bigg( \sum_{l=1}^{|\mathcal{P}|} {w^{(l)}_u \sigma\big(\mathbf{M}^{(l)} \bm{e}^{(l)}_u+\bm{b}^{(l)}\big)}\bigg),
\end{equation}
where $\sigma(\cdot)$ is a non-linear function, \ie sigmoid function in our work. Although we only use two non-linear transformations, it is flexible to extend to multiple non-linear layers, \eg Multi-Layer Perceptrons.
\end{itemize}

\subsubsection{Model Learning}
We blend the fusion function into matrix factorization framework for learning the parameters of the proposed model.
The objective can be formulated as follows:

\begin{align}
 \pounds &= \sum_{\langle u, i, r_{u,i}\rangle \in \mathcal{R}}{(r_{u,i} - \widehat{r_{u,i}})}^2  + \lambda \sum_u{(\|\mathbf{x}_u\|_2 + \|\mathbf{y}_i\|_2} \nonumber \\
 &+ \|\bm{\gamma}^{(U)}_u\|_2 + \|\bm{\gamma}^{(I)}_i\|_2 + \|\bm{\Theta}^{(U)}\|_2 + \|\bm{\Theta}^{(I)}\|_2),
\end{align}

\noindent where $\widehat{r_{u,i}}$ is the predicted rating using Eq.~\ref{eq-predictor} by the proposed model, $\lambda$ is the regularization parameter,
and $\bm{\Theta}^{(U)}$ and $\bm{\Theta}^{(I)}$ are the parameters of the function $g(\cdot)$ for users and items respectively. We adopt SGD to efficiently optimize the final objective.
The update of original latent factors $\{\mathbf{x}_u\}$ and $\{\mathbf{y}_i\}$ are the same as that of standard MF in Eq.~\ref{eq-mf}.
The parameters of the proposed model will be updated as follows:

\begin{small}
\begin{eqnarray}
\bm{\Theta}_{u,l}^{(U)} &\leftarrow& \bm{\Theta}_{u,l}^{(U)} - \eta \cdot (-\alpha (r_{u,i}-\widehat{r_{u,i}})\bm{\gamma}_i^{(I)}\frac{\partial{\bm{e}_{u}^{(U)}}}{\partial{\bm{\Theta}_{u,l}^{(U)}}} + {\lambda_{\Theta}}\bm{\Theta}_{u,l}^{(U)}), \nonumber \\
\\
\bm{\gamma}_u^{(U)} &\leftarrow& \bm{\gamma}_u^{(U)} - \eta \cdot (-\beta (r_{u,i}-\widehat{r_{u,i}})\bm{e}_i^{(I)} + \lambda_{\gamma}\bm{\gamma}_u^{(U)}),\\
\nonumber\\
\bm{\Theta}_{i,l}^{(I)} &\leftarrow& \bm{\Theta}_{i,l}^{(I)} - \eta \cdot (-\beta (r_{u,i}-\widehat{r_{u,i}})\bm{\gamma}_u^{(U)}\frac{\partial{\bm{e}_{i}^{(I)}}}{\partial{\bm{\Theta}_{i,l}^{(I)}}} + {\lambda_{\Theta}}\bm{\Theta}_{i,l}^{(I)}), \nonumber \\
\\
\bm{\gamma}_i^{(I)} &\leftarrow& \bm{\gamma}_i^{(I)} - \eta \cdot (-\alpha (r_{u,i}-\widehat{r_{u,i}})\bm{e}_u^{(U)} + \lambda_{\gamma}\bm{\gamma}_i^{(I)}),
\end{eqnarray}
\end{small}
where $\eta$ is the learning rate, $\lambda_{\Theta}$ is the regularization for parameters $\bm{\Theta}^{(U)}$ and $\bm{\Theta}^{(I)}$,  and $\lambda_{\gamma}$ is the regularization for parameters $\bm{\gamma}^{(U)}$ and $\bm{\gamma}^{(I)}$.  In our work, we utilize the sigmod function for non-linear transformation, and we can take advantage of the properties of sigmod function for ease of derivative calculation. It is worth noting that the symbol $\bm{\Theta}$ denotes all the parameters in the fusion function, and the calculation of $\frac{\partial{\bm{e}_{i}}}{\partial{\bm{\Theta}_{i,l}}}$ will be different for different parameters in $\bm{\Theta}$. Next, we present the detailed derivation of $\frac{\partial{\bm{e}_{i}}}{\partial{\bm{\Theta}_{i,l}}}$ for personalized non-linear fusion function

\begin{eqnarray}\label{eq-gradient}
\small
&&\frac{\partial{\bm{e}_{i}}}{\partial{\bm{\Theta}_{i,l}}}=\\
&&\begin{cases}
w^{(l)}_i\sigma(Z_s)\sigma(Z_f)(1-\sigma(Z_s))(1-\sigma(Z_f))e^{(l)}_i, &\text{$\bm{\Theta} = \bm{M}$};\nonumber\\
\\
w^{(l)}_i\sigma(Z_s)\sigma(Z_f)(1-\sigma(Z_s))(1-\sigma(Z_f)), &\text{$\bm{\Theta} = \bm{b}$}; \nonumber\\
\\
\sigma(Z_s)\sigma(Z_f)(1-\sigma(Z_s)), &\text{$\bm{\Theta} = w$},
\end{cases}
\end{eqnarray}
where $Z_s = \sum_{l=1}^{|\mathcal{P}|} {w^{(l)}_i \sigma\big(\mathbf{M}^{(l)} \bm{e}^{(l)}_i+\bm{b}^{(l)}\big)}$ and $Z_f = \mathbf{M}^{(l)} \bm{e}^{(l)}_i+\bm{b}^{(l)}$. The derivations of $\frac{\partial{\bm{e}_{i}}}{\partial{\bm{\Theta}_{i,l}}}$ can be calculated in the above way for both users and items. We omit the derivations for linear fusion functions, since it is relatively straightforward.
The whole algorithm framework is shown in Algorithm~\ref{alg_herec}.
In lines 1-6, we perform HIN embedding to obtain the representations of users and items. And in lines 8-19, we adopt the SGD algorithm to optimize the parameters in the fusion function and rating function.

%\begin{eqnarray}\label{eq-rw}
%&&P({n_{t+1}=x |n_{t}=v, \rho})\\
%&=&\begin{cases}
%\frac{1}{|\mathcal{N}^{A_{t+1}}(v)|}, &\text{($v, x$) $\in$ $\mathcal{E}$ and $\phi(x) = A_{t+1}$};\nonumber\\
%0,& \text{otherwise},
%\end{cases}
%\end{eqnarray}

%We omit the learning of $\bm{\Theta}_{i,l}^{(I)}$ and $\bm{\gamma}_i^{(I)}$ due to the similar derivations.

\begin{algorithm}[htb]
\caption{The overall learning algorithm of HERec.}
\label{alg_herec}
\begin{algorithmic}[1]
\Require

the rating matrix $\mathcal{R}$;
%The final HIN embeddings of users, items, $\bm{e}^{(U)}, \bm{e}^{(V)}$;
the learning rate $\eta$;
the adjustable parameters $\alpha, \beta$;
the regularization parameter $\lambda$;
the meta-path sets for users and items, $\mathcal{P}^{(U)}$ and $\mathcal{P}^{(I)}$.
\Ensure
the latent factors for users and items, $\mathbf{x}$ and $\mathbf{y}$;
the latent factors to pair HIN embedding of users and items, $\bm{{\gamma}}^{(U)}$ and $\bm{{\gamma}}^{(I)}$;
the parameters of the fusion function for users and items, $\bm{\Theta}^{(U)}$  and $\bm{\Theta}^{(I)}$
\For {$l = 1$ to $|\mathcal{P}^{(U)}|$}
    \State Obtain users' embeddings $\{\bm{e}^{(l)}_u\}$  based on meta-path $\mathcal{P}^{(U)}_l$ according to Algorithm \ref{alg_embedding};
\EndFor
\For {$l = 1$ to $|\mathcal{P}^{(I)}|$}
    \State Obtain items' embeddings $\{\bm{e}^{(l)}_i\}$ based on the meta-path set $\mathcal{P}^{(I)}_l$  according to Algorithm \ref{alg_embedding};
\EndFor
\State Initialize $\mathbf{x}, \mathbf{y}, \bm{{\gamma}}^{(U)}, \bm{{\gamma}}^{(I)}, \bm{\Theta}^{(U)}, \bm{\Theta}^{(I)}$ by standard normal distribution;
\While {not convergence}
    \State Randomly select a triple $\langle u, i, r_{u,i}\rangle \in \mathcal{R}$;
    \State Update  $\mathbf{x}_u, \mathbf{y}_i$ by typical MF;
    \For {$l = 1$ to $|\mathcal{P}^{(U)}|$}
        \State Calculate $\frac{\partial{\bm{e}_{u}^{(U)}}}{\partial{\bm{\Theta}_{u,l}^{(U)}}}$ by Eq. 14;
        \State Update  $\bm{\Theta}_{u,l}^{(U)}$ by Eq. 10;
    \EndFor
    \State Update $\bm{\gamma}_u^{(U)}$ by Eq. 11;
     \For {$l = 1$ to $|\mathcal{P}^{(I)}|$}
        \State Calculate $\frac{\partial{\bm{e}_{i}^{(I)}}}{\partial{\bm{\Theta}_{i,l}^{(I)}}}$ by Eq. 14;
        \State Update  $\bm{\Theta}_{i,l}^{(I)}$ by Eq. 12;
    \EndFor
    \State Update $\bm{\gamma}_i^{(I)}$ by Eq. 13;
\EndWhile
\State \Return {$\mathbf{x}, \mathbf{y}, \bm{{\gamma}}^{(U)}, \bm{{\gamma}}^{(I)}, \bm{\Theta}^{(U)}, \bm{\Theta}^{(I)}$}.
\end{algorithmic}
\end{algorithm}

\subsubsection{Complexity Analysis}
HERec contains two major parts: (1) HIN embedding. The complexity of deepwalk is $\mathcal{O}(d \cdot |V|)$, where $d$ is the embedding dimension and $|V|$ is the number of nodes in the network. Therefore it takes $\mathcal{O}(d \cdot |\mathcal{U}|)$ and $\mathcal{O}(d \cdot |\mathcal{I}|)$ to learn users' and items' embeddings according to a single meta-path, respectively. And the total complexity of HIN embedding is $\mathcal{O}(|\mathcal{P}|\cdot d \cdot (|\mathcal{U}|+|\mathcal{I}|))$ since the number of selected meta-paths is $|\mathcal{P}|$. It is worth noting that HIN embedding can be easily trained in parallel and we will implement it using a multi-thread mode in order to improve the efficiency of model. (2) Matrix factorization. For each triplet $\langle u, i, r_{u,i}\rangle$, updating $\bm{x}_u$, $\bm{y}_i$, $\bm{\gamma}_u^{(U)}$, $\bm{\gamma}_i^{(I)}$ takes $\mathcal{O}(D)$ time, where $D$ is the number of latent factors. And  updating $\bm{\Theta}_u^{(U)}$, $\bm{\Theta}_i^{(I)}$ takes $\mathcal{O}(|\mathcal{P}| \cdot D \cdot d)$ time to learn the transformation matrices  $\mathbf{M}$ for all meta-paths.
In the proposed approach, $|\mathcal{P}|$ is generally small, and $d$ and $D$ are at most several hundreds, which makes the proposed method efficient in large datasets. Specially, SGD has very good practice performance, and we have found that it has a fast convergence rate on our datasets.

\section{Experiments \label{sec-exp}}
%In this section, we will verify the superiority of HERec by performing experiments on three real datasets and comparing it to the state-of-the-arts.
%In this section, we construct the experimental evaluation and present the  result analysis.
In this section, we will demonstrate the effectiveness of HERec by performing experiments on three real datasets compared to the state-of-the-art recommendation methods.

\begin{table*}[t]%[htbp]
\centering
%\scriptsize
\caption{\label{tab_Data} Statistics of the three datasets.}
{
\begin{tabular}{|c||c|c|c|c|c|c|}
\hline
{Dataset} & {Relations} & {Number} & {Number} & {Number} & {Ave. degrees} & {Ave. degrees}\\
{(Density)} & {(A-B)} & {of A} & {of B} & {of (A-B)}  & {of A} & {of B}\\
\hline
\hline
\multirow{6}{*}{Douban Movie} & {User-Movie} & {13,367} & {12,677} & {1,068,278} & {79.9} & {84.3}  \\
\cline{2-7}
\multirow{6}{*}{} &  {User-User} & {2,440} & {2,294} & {4,085} & {1.7} & {1.8}    \\
\cline{2-7}
\multirow{6}{*}{(0.63\%)} &{User-Group} & {13,337} & {2,753} & {570,047} & {42.7} & {207.1}\\
\cline{2-7}
\multirow{6}{*}{} & {Movie-Director} & {10,179} & {2,449} & {11,276} & {1.1} & {4.6}  \\
\cline{2-7}
\multirow{6}{*}{} & {Movie-Actor} & {11,718} & {6,311} & {33,587} & {2.9} & {5.3}  \\
\cline{2-7}
\multirow{6}{*}{} & {Movie-Type} & {12,678} & {38} & {27,668} & {2.2} & {728.1}  \\
\hline
\hline
\multirow{5}{*}{Douban Book} & {User-Book} & {13,024} & {22,347} & {792,026} & {60.8} & {35.4}  \\
\cline{2-7}
\multirow{5}{*}{} & {User-User} & {12,748} & {12,748} & {169,150} & {13.3} & {13.3}  \\
\cline{2-7}
\multirow{5}{*}{(0.27\%)} & {Book-Author} & {21,907} & {10,805} & {21,905} & {1.0} & {2.0}  \\
\cline{2-7}
\multirow{5}{*}{} & {Book-Publisher} & {21,773} & {1,815} & {21,773} & {1.0} & {11.9}  \\
\cline{2-7}
\multirow{5}{*}{} & {Book-Year} & {21,192} & {64} & {21,192} & {1.0} & {331.1}  \\
\hline
\hline
\multirow{5}{*}{Yelp} & {User-Business} & {16,239} & {14,284} & {198,397} & {12.2} & {13.9}  \\
\cline{2-7}
\multirow{5}{*}{(0.08\%)} & {User-User} & {10,580} & {10,580} & {158,590} & {15.0} & {15.0}  \\
\cline{2-7}
\multirow{5}{*}{} & {User-Compliment} & {14,411} & {11} & {76,875} & {5.3} & {6988.6}  \\
\cline{2-7}
\multirow{5}{*}{} & {Business-City} & {14,267} & {47} & {14,267} & {1.0} & {303.6}  \\
\cline{2-7}
\multirow{5}{*}{} & {Business-Category} & {14,180} & {511} & {40,009} & {2.8} & {78.3}  \\
\hline

\end{tabular}}
\end{table*}

\begin{table}[t]%[htbp]
\center
%\scriptsize
\caption{\label{tab_metapath} The selected meta-paths for  three datasets in our work.}
{\begin{tabular}{|c||c|}
\hline
{Dataset} & {Meta-paths}\\
\hline\hline
\multirow{2}{*}{Douban Movie} & {UMU, UMDMU, UMAMU, UMTMU}\\
\multirow{2}{*}{} & {MUM, MAM, MDM, MTM} \\
\hline
\multirow{2}{*}{Douban Book} & {UBU, UBABU, UBPBU, UBYBU} \\
\multirow{2}{*}{} & {BUB, BPB, BYB} \\
\hline
\multirow{2}{*}{Yelp} & {UBU, UBCiBU, UBCaBU} \\
\multirow{2}{*}{} & {BUB, BCiB, BCaB} \\
\hline
\end{tabular}}
\end{table}

\begin{table*}[t]
\centering
%\scriptsize
\caption{\label{tab_Effectiveness} Results of effectiveness experiments on three datasets. A smaller MAE or RMSE value indicates a better performance. For ease of reading the results, we also report the improvement w.r.t. the PMF model for each other method. A larger improvement ratio indicates a better performance. }
{
\begin{tabular}{|c|c|c||c|c|c|c|c|c||c|c||c|}
\hline
{Dataset} & {Training} & {Metrics}  & {PMF} &{SoMF} & {FM$_{HIN}$} & {HeteMF}& {SemRec}  & {DSR} & {HERec$_{dw}$} & {HERec$_{mp}$}&{HERec}\\
\hline
\hline
\multirow{16}{*}{Douban}& \multirow{4}{*}{80\%} & {MAE} & {0.5741} & {0.5817} & {0.5696} & {0.5750} & {0.5695} &  {0.5681}  & {0.5703} & {\textbf{0.5515}} &{0.5519} \\
\multirow{16}{*}{} &\multirow{4}{*}{}& {Improve} & {} & {-1.32\%} & {+0.78\%} & {-0.16\%} & {+0.80\%}  & {+1.04\%} & {+0.66\%} &{+3.93\%} & {+3.86\%} \\
\cline{3-12}
\multirow{16}{*}{Movie} &\multirow{4}{*}{}& {RMSE} & {0.7641} & {0.7680} &   {0.7248} & {0.7556}& {0.7399} & {0.7225} & {0.7446} & {0.7121} & {\textbf{0.7053}} \\
\multirow{16}{*}{} &\multirow{4}{*}{}& {Improve} & {} & {-0.07\%} & {+5.55\%} & {+1.53\%} & {+3.58\%}  & {+5.85\%} & {+2.97\%} &{+7.20\%} & {+8.09\%} \\
\cline{2-12}
\multirow{16}{*}{} &\multirow{4}{*}{}\multirow{4}{*}{60\%} & {MAE} & {0.5867} & {0.5991} & {0.5769} & {0.5894} & {0.5738} &  {0.5831} & {0.5838} & {0.5611} &{\textbf{0.5587}} \\
\multirow{16}{*}{} &\multirow{4}{*}{}& {Improve} & {} & {-2.11\%} & {+1.67\%} & {-0.46\%} & {+2.19\%}  & {+0.61\%} & {+0.49\%} &{+4.36\%} & {+4.77\%} \\
\cline{3-12}
\multirow{16}{*}{} &\multirow{4}{*}{}& {RMSE} & {0.7891} & {0.7950} & {0.7842} & {0.7785} & {0.7551} &  {0.7408} & {0.7670} & {0.7264} &{\textbf{0.7148}} \\
\multirow{16}{*}{} &\multirow{4}{*}{}& {Improve} & {} & {-0.75\%} & {+0.62\%} & {+1.34\%} & {+4.30\%}  & {+6.12\%} & {+2.80\%} &{+7.94\%} & {+9.41\%} \\
\cline{2-12}
\multirow{16}{*}{} &\multirow{4}{*}{}\multirow{4}{*}{40\%} & {MAE} & {0.6078} & {0.6328} &  {0.5871} & {0.6165} & {0.5945} &  {0.6170} & {0.6073} & {0.5747} &{\textbf{0.5699}}\\
\multirow{16}{*}{} &\multirow{4}{*}{}& {Improve} & {} & {-4.11\%} & {+3.40\%} & {-1.43\%} & {+2.18\%}  & {-1.51\%} & {+0.08\%} &{+5.44\%} & {+6.23\%} \\
\cline{3-12}
\multirow{16}{*}{} &\multirow{4}{*}{}& {RMSE} & {0.8321} & {0.8479} & {0.7563} & {0.8221} & {0.7836} &  {0.7850} & {0.8057} & {0.7429} &{\textbf{0.7315}} \\
\multirow{16}{*}{} &\multirow{4}{*}{}& {Improve} & {} & {-1.89\%} & {+9.10\%} & {+1.20\%} & {+5.82\%}  & {+5.66\%} & {+3.17\%} &{+10.71\%} & {+12.09\%} \\
\cline{2-12}
\multirow{16}{*}{} &\multirow{4}{*}{}\multirow{4}{*}{20\%} & {MAE}& {0.7247}& {0.6979} & {0.6080} & {0.6896} & {0.6392} &  {0.6584} & {0.6699} & {0.6063} &{\textbf{0.5900}} \\
\multirow{16}{*}{} &\multirow{4}{*}{}& {Improve} & {} & {+3.69\%} & {+16.10\%} & {+4.84\%} & {+11.79\%}  & {+9.14\%} & {+7.56\%} &{+16.33\%} & {+18.59\%} \\
\cline{3-12}
\multirow{16}{*}{} &\multirow{4}{*}{}& {RMSE} & {0.9440} & {0.9852} & {0.7878} & {0.9357} & {0.8599} &  {0.8345} & {0.9076} & {0.7877}  &{\textbf{0.7660}}\\
\multirow{16}{*}{} &\multirow{4}{*}{}& {Improve} & {} & {-4.36\%} & {+16.55\%} & {+0.88\%} & {+8.91\%}  & {+11.60\%} & {+3.86\%} &{+16.56\%}& {+18.86\%} \\

\hline
\hline
\multirow{16}{*}{Douban}& \multirow{4}{*}{80\%} & {MAE} & {0.5774} & {0.5756}  & {0.5716} & {0.5740}   & {0.5675}   &  {0.5740}    & {0.5875}   & {0.5591} &{\textbf{0.5502}} \\
\multirow{16}{*}{} &\multirow{4}{*}{}& {Improve} & {} & {+0.31\%} & {+1.00\%} & {+0.59\%} & {+1.71\%}  & {+0.59\%} & {-1.75\%} &{+3.17\%} & {+4.71\%} \\
\cline{3-12}
\multirow{16}{*}{Book} &\multirow{4}{*}{}& {RMSE}             & {0.7414} & {0.7302}   & {0.7199} & {0.7360}   & {0.7283}   &  {0.7206}    & {0.7450}  & {0.7081} &{\textbf{0.6811}}\\
\multirow{16}{*}{} &\multirow{4}{*}{}& {Improve} & {} & {+1.55\%} & {+2.94\%} & {+0.77\%} & {+1.81\%}  & {+2.84\%} & {-0.44\%} &{+4.53\%} & {+8.17\%} \\
\cline{2-12}
\multirow{16}{*}{} &\multirow{4}{*}{}\multirow{4}{*}{60\%} & {MAE} & {0.6065} & {0.5903}   &{0.5812} & {0.5823}   & {0.5833}   &  {0.6020}   & {0.6203}  & {0.5666} &{\textbf{0.5600}}\\
\multirow{16}{*}{} &\multirow{4}{*}{}& {Improve} & {} & {+2.67\%} & {+4.17\%} & {+3.99\%} & {+3.83\%}  & {+0.74\%} & {-2.28\%} &{+6.58\%} & {+7.67\%} \\
\cline{3-12}
\multirow{16}{*}{} &\multirow{4}{*}{}& {RMSE}    & {0.7908} & {0.7518}   &{0.7319} & {0.7466}   & {0.7505}   &  {0.7552}   & {0.7905}  & {0.7318} &{\textbf{0.7123}}\\
\multirow{16}{*}{} &\multirow{4}{*}{}& {Improve} & {} & {+4.93\%} & {+7.45\%} & {+5.59\%} & {+5.10\%}  & {+4.50\%} & {+0.04\%} &{+7.46\%} & {+9.93\%} \\
\cline{2-12}
\multirow{16}{*}{} &\multirow{4}{*}{}\multirow{4}{*}{40\%} & {MAE} & {0.6800} & {0.6161}   &{0.6028} & {0.5982}   & {0.6025}   &  {0.6271}   & {0.6976}  & {0.5954} &{\textbf{0.5774}}\\
\multirow{16}{*}{} &\multirow{4}{*}{}& {Improve} & {} & {+9.40\%} & {+11.35\%} & {+12.03\%} & {+11.40\%}  & {+7.78\%} & {-2.59\%} &{+12.44\%}& {+15.09\%} \\
\cline{3-12}
\multirow{16}{*}{} &\multirow{4}{*}{}& {RMSE}    & {0.9203} & {0.7936}   &{0.7617} & {0.7779}   & {0.7751}   &  {0.7730}   & {0.9022}  & {0.7703} &{\textbf{0.7400}}\\
\multirow{16}{*}{} &\multirow{4}{*}{}& {Improve} & {} & {+13.77\%} & {+17.23\%} & {+15.47\%} & {+15.78\%}  & {+16.01\%} & {+1.97\%} &{+16.30\%} & {+19.59\%} \\
\cline{2-12}
\multirow{16}{*}{} &\multirow{4}{*}{}\multirow{4}{*}{20\%} & {MAE}& {1.0344} & {0.6327}   &{0.6396} & {0.6311}   & {0.6481}   &  {0.6300}   & {1.0166}   & {0.6785}  &{\textbf{0.6450}}\\
\multirow{16}{*}{} &\multirow{4}{*}{}& {Improve} & {} & {+38.83\%} & {+38.17\%} & {+38.99\%} & {+37.35\%}  & {+39.10\%} & {+1.72\%} &{+34.41\%} & {+37.65\%} \\
\cline{3-12}
\multirow{16}{*}{} &\multirow{4}{*}{}& {RMSE}    & {1.4414} & {0.8236}   &{0.8188} & {0.8304}   & {0.8350}   &  {0.8200}   & {1.3205}   & {0.8869} &{\textbf{0.8581}} \\
\multirow{16}{*}{} &\multirow{4}{*}{}& {Improve} & {} & {+42.86\%} & {+43.19\%} & {+42.39\%} & {+42.07\%}  & {+43.11\%} & {+8.39\%} &{+38.47\%} & {+40.47\%} \\
%\cline{2-12}
%{} &\multicolumn{2}{c||}{Average Rank} & {8.88} & {5.88} & {5.50} & {5.53} & {4.88} & {8.75} & {2.25} & {2.37} & {\textbf{1.63}}\\
\hline
\hline
\multirow{16}{*}{Yelp}& \multirow{4}{*}{90\%} & {MAE} & {1.0412} & {1.0095}   &{0.9013} & {0.9487}   & {0.9043}   &  {0.9054}    & {1.0388}   & {0.8822} &{\textbf{0.8395}} \\
\multirow{16}{*}{} &\multirow{4}{*}{}& {Improve} & {} & {+3.04\%} & {+13.44\%} & {+8.88\%} & {+13.15\%}  & {+13.04\%} & {+0.23\%} &{+15.27\%} & {+19.37\%} \\
\cline{3-12}
\multirow{16}{*}{} &\multirow{4}{*}{}& {RMSE}             & {1.4268} & {1.3392}   &{1.1417} & {1.2549}   & {1.1637}   &  {1.1186}    & {1.3581}  & {1.1309}  &{\textbf{1.0907}}\\
\multirow{16}{*}{} &\multirow{4}{*}{}& {Improve} & {} & {+6.14\%} & {+19.98\%} & {+12.05\%} & {+18.44\%}  & {+21.60\%} & {+4.81\%} &{+20.74\%} & {+23.56\%} \\
\cline{2-12}
\multirow{16}{*}{} &\multirow{4}{*}{}\multirow{4}{*}{80\%} & {MAE} & {1.0791} & {1.0373}   &{0.9038} & {0.9654}   & {0.9176}   &  {0.9098}   & {1.0750}   & {0.8953}  &{\textbf{0.8475}} \\
\multirow{16}{*}{} &\multirow{4}{*}{}& {Improve} & {} & {+3.87\%} & {+16.25\%} & {+10.54\%} & {+14.97\%}  & {+15.69\%} & {+0.38\%} &{+17.03\%} & {+21.46\%} \\
\cline{3-12}
\multirow{16}{*}{} &\multirow{4}{*}{}& {RMSE}    & {1.4816} & {1.3782}   &{1.1497} & {1.2799}   & {1.1771}   &  {1.1208}   & {1.4075}   & {1.1516} &{\textbf{1.1117}}\\
\multirow{16}{*}{} &\multirow{4}{*}{}& {Improve} & {} & {+6.98\%} & {+22.40\%} & {+13.61\%} & {+20.55\%}  & {+24.35\%} & {+5.00\%} &{+22.27\%} & {+24.97\%} \\
\cline{2-12}
\multirow{16}{*}{} &\multirow{4}{*}{}\multirow{4}{*}{70\%} & {MAE} & {1.1170} & {1.0694}   &{0.9108} & {0.9975}   & {0.9407}   &  {0.9429}   & {1.1196}  & {0.9043} &{\textbf{0.8580}}\\
\multirow{16}{*}{} &\multirow{4}{*}{}& {Improve} & {} & {+4.26\%} & {+18.46\%} & {+10.70\%} & {+15.78\%}  & {+15.59\%} & {-0.23\%} &{+19.04\%} & {+23.19\%} \\
\cline{3-12}
\multirow{16}{*}{} &\multirow{4}{*}{}& {RMSE}    & {1.5387} & {1.4201}   &{1.1651} & {1.3229}   & {1.2108}   &  {1.1582}   & {1.4632}  & {1.1639}  &{\textbf{1.1256}}\\
\multirow{16}{*}{} &\multirow{4}{*}{}& {Improve} & {} & {+7.71\%} & {+24.28\%} & {+14.02\%} & {+21.31\%}  & {+24.73\%} & {+4.91\%} &{+24.36\%} & {+26.85\%} \\
\cline{2-12}
\multirow{16}{*}{} &\multirow{4}{*}{}\multirow{4}{*}{60\%} & {MAE}& {1.1778} & {1.1135}   &{0.9435} & {1.0368}   & {0.9637}   &  {1.0043}   & {1.1691}  & {0.9257}  &{\textbf{0.8759}}\\
\multirow{16}{*}{} &\multirow{4}{*}{}& {Improve} & {} & {+5.46\%} & {+19.89\%} & {+11.97\%} & {+18.18\%}  & {+14.73\%} & {+0.74\%} &{+21.40\%} & {+25.63\%} \\
\cline{3-12}
\multirow{16}{*}{} &\multirow{4}{*}{}& {RMSE}    & {1.6167} & {1.4748}   &{1.2039} & {1.3713}   & {1.2380}   &  {1.2257}   & {1.5182}   & {1.1887}  &{\textbf{1.1488}}\\
\multirow{16}{*}{} &\multirow{4}{*}{}& {Improve} & {} & {+8.78\%} & {+25.53\%} & {+15.18\%} & {+23.42\%}  & {+24.19\%} & {+6.09\%} &{+26.47\%} & {+28.94\%} \\
%\cline{2-12}
%{} &\multicolumn{2}{c||}{Average Rank} & {9.00} & {7.88} & {6.00} & {4.63} & {4.38} & {7.13} & {3.00} & {1.75} & {\textbf{1.25}}\\
\hline
%\hline
%\multicolumn{3}{|c||}{Average Rank} & {10.33} & {8.79} &{5.25} & {7.83} & {6.54} &  {6.13} & {9.5} & {4.17} & {3.2} & {2.4} & {\textbf{1.79}}\\
%\hline
\end{tabular}}
\end{table*}

\subsection{Evaluation Datasets}
We adopt three widely used datasets from different domains, consisting of Douban Movie dataset \footnote{http://movie.douban.com} from movie domain, Douban Book dataset \footnote{http://book.douban.com} from book domain, and Yelp dataset \footnote{http://www.yelp.com/dataset-challenge} from business domain. Douban Movie dataset includes 13,367 users and 12,677 movies with 1,068,278 movie ratings ranging from 1 to 5. Moreover, the dataset also includes social relations and attribute information of users and movies. As for the Douban Book dataset, it includes 13,024 users and 22,347 books with 792,026 ratings ranging from 1 to 5. Yelp dataset records user ratings on local businesses and contains social relations and attribute information of businesses, which includes 16,239 users and 14,282 local businesses with 198,397 ratings raging from 1 to 5. The detailed descriptions of the three datasets are shown in Table~\ref{tab_Data}, and the network schemas of the three datasets are plotted in Fig.~\ref{fig_schema}. Besides the domain variation, these three datasets also have different rating sparsity degrees: the Yelp dataset is very sparse, while the Douban Movie dataset is much denser.

\subsection{Metrics}
We use the widely used mean absolute error (MAE) and root mean square error (RMSE) to measure the quality of recommendation of different models. The metrics MAE and RMSE are defined as follows:

\begin{equation}
MAE = \frac{1}{|\mathcal{D}_{test}|}\sum_{(i, j) \in \mathcal{D}_{test}}{|r_{i,j} - \widehat{r_{i,j}}|},
\end{equation}
\begin{equation}
RMSE = \sqrt{\frac{1}{|\mathcal{D}_{test}|}\sum_{(i, j) \in \mathcal{D}_{test}}{(r_{i,j} - \widehat{r_{i,j}})^2}},
\end{equation}
where $r_{i,j}$ is the actual rating that user $i$ assigns to item $j$, $\widehat{r_{i,j}}$ is the predicted rating from a model, and $\mathcal{D}_{test}$ denotes the test set of rating records. From the definition, we can see that a smaller MAE or RMSE value indicates a better performance.

%\subsection{Evaluation Datasets and Metrics}
%
%We adopt three widely used datasets from different domains, including Douban Movie dataset \footnote{http://movie.douban.com}, Douban Book dataset \footnote{http://book.douban.com}, and Yelp dataset \footnote{http://www.yelp.com/dataset-challenge}. The detailed statistics of three datasets are
%summarized in Table~\ref{tab_Data}.
%We use two commonly adopted metrics to measure the performance of rating prediction, namely mean absolute error (MAE) and root mean square error (RMSE) \cite{willmott2005advantages}.

\subsection{Methods to Compare}
%In order to verify the effectiveness of HERec, we compare HERec with quite a few baselines, some of which represent state-of-the-art for HIN-based recommendation. Since there are three embedding fusion functions in HERec, we include three versions of HERec: HERec$_{sl}$, HERec$_{pl}$, and HERec$_{pnL}$. Moreover, we also design two methods, called HERec$_{dw}$ and HERec$_{mp}$, which replace our HIN embedding with two well known network embeddings under the framework of HERec. As the baselines, six representative rating predication methods are employed, including two classical matrix factorization methods (i.e., PMF and SoMF) and  four HIN based recommendation methods (FM$_{HIN}$, HeteMF, SemRec and DSR).
We consider the following methods to compare:

\textbullet\ \textbf{PMF}~\cite{mnih2008probabilistic}: It is the classic probabilistic matrix factorization model by explicitly factorizing the rating matrix into two low-dimensional matrices.

\textbullet\ \textbf{SoMF}~\cite{ma2011recommender}: It incorporates social network information into the recommendation model.
The social relations are characterized by a social regularization term, and integrated into the basic matrix factorization model.

\textbullet\ \textbf{FM$_{HIN}$}: It is the context-aware factorization machine~\cite{rendle2012factorization}, which is able to utilize various kinds of auxiliary information. In our experiments, we extract heterogeneous information as context features and incorporate them into the factorization machine for rating prediction. We implement the model with the code from the authors~\footnote{http://www.libfm.org/}.

\textbullet\ \textbf{HeteMF}~\cite{yu2013collaborative}: It is a MF based recommendation method, which utilizes entity similarities calculated over HINs using meta-path based algorithms.

\textbullet\ \textbf{SemRec}~\cite{shi2015semantic}: It is a collaborative filtering method on weighted heterogeneous information network, which is constructed by connecting users and items with the same ratings. It flexibly integrates heterogeneous information for recommendation by weighted meta-paths and a weight ensemble method. We implement the model with the code from the authors~\footnote{https://github.com/zzqsmall/SemRec}.

\textbullet\ \textbf{DSR}~\cite{zheng2017recommendation}: It is a MF based recommendation method with dual similarity regularization, which imposes the constraint on users and items with high and low similarities simultaneously.

\textbullet \textbf{HERec$_{dw}$}: It is a variant of HERec, which adopts the homogeneous network embedding method deepwalk~\cite{perozzi2014deepwalk} and ignores the heterogeneity
of nodes in HINs. We use the code of deepwalk provided by the authors~\footnote{https://github.com/phanein/deepwalk}.

\textbullet \textbf{HERec$_{mp}$}: It can be view as a variant of HERec, which combines the heterogeneous network embedding of metapath2vec++~\cite{dong2017metapath2vec} and still preserves the superiority of HINs. We use the code of metapath2vec++ provided by the authors~\footnote{https://ericdongyx.github.io/metapath2vec/m2v.html}.

%\textbullet\ \textbf{HERec$_{dw}$}, \textbf{HERec$_{mp}$}: These two HERec variants adopt homogeneous network embedding method deepwalk~\cite{perozzi2014deepwalk} and heterogeneous network embedding method  metapath2vec++~\cite{dong2017metapath2vec}, respectively.

%\textbullet\ \textbf{HERec$_{sl}$}: It is a version of HERec correspond to the simple linear fusion function in Eq.~\ref{eq-slf}.

%\textbullet\ \textbf{HERec$_{pl}$}: It is a version of HERec correspond to the personalized linear fusion function in Eq.~\ref{eq-plf}.

%\textbullet\ \textbf{HERec$_{pnl}$}: It is a version of HERec correspond to the personalized non-linear fusion function in Eq.~\ref{eq-pnlf}.
%\textbullet\ \textbf{HERec$_{pnl}$}: These three implementations of HERec correspond to the three fusion functions in Eq.~\ref{eq-slf}, Eq.~\ref{eq-plf} and Eq.~\ref{eq-pnlf}, namely simple linear fusion, personalized linear fusion and personalized non-linear fusion.
\textbullet\ \textbf{HERec}: It is the proposed recommendation model based on heterogeneous information network embedding. In the effectiveness experiments, we utilize the personalized non-linear fusion function, which is formulated as Eq.~\ref{eq-pnlf}. And the performance of different fusion function will be reported in the later section.

The selected baselines have a comprehensive coverage of existing rating prediction methods. PMF and SoMF are classic MF based rating prediction methods; FM$_{HIN}$, HeteMF, SemRec and DSR are HIN based recommendation methods using meta-path based similarities, which represent state-of-the-art HIN based methods. The proposed approach contains an HIN embedding component, which can be replaced by other network embedding methods. Hence, two variants HERec$_{dw}$ and HERec$_{mp}$
are adopted as baselines.

Among these methods, HIN based methods need to specify the used meta-paths. We present the used meta-paths  in Table~\ref{tab_metapath}.
Following~\cite{sun2011pathsim}, we only select short meta-paths of at most four steps, since long meta-paths are likely to introduce noisy semantics.
We set parameters for HERec as follows: the embedding dimension number $d = 64$ and the tuning coefficients $\alpha = \beta = 1.0$.
Following~\cite{perozzi2014deepwalk}, the path length for random walk is set to 40.
For HERec$_{dw}$ and HERec$_{mp}$, we set the embedding dimension number $d = 64$ for fairness.
Following~\cite{shi2016integrating,zheng2017recommendation}, we set the number of latent factors for all the MF based methods to 10.
Other baseline parameters either adopt the original optimal setting or are optimized by using a validation set of 10\% training data.

\subsection{Effectiveness Experiments}
For each dataset, we split the entire rating records into a training set and a test set. %We train the model using the training set and compare the performance on the test set.
For Double Movie and Book datasets, we set four training ratios as in $\{80\%, 60\%, 40\%, 20\%\}$; while for Yelp dataset, following~\cite{shi2015semantic}, we set four larger training ratios as in $\{90\%, 80\%, 70\%, 60\%\}$, since
it is too sparse. For each ratio, we randomly generate ten evaluation sets. We average the  results as the final performance.
The results are shown in Table \ref{tab_Effectiveness}. The major findings from the experimental results are summarized as follows:

(1) Among these baselines, HIN based methods (HeteMF, SemRec, FM$_{HIN}$ and DSR) perform better than traditional MF based methods (PMF and SoMF), which indicates the usefulness of the heterogeneous information. It is worthwhile to note that the FM$_{HIN}$ model works very well among the three HIN based baselines. An intuitive explanation is that in our datasets, most of the original features are the attribute information of users or items, which are likely to contain useful evidence to improve the recommendation performance.

(2) The proposed HERec method (corresponding to the last column) is consistently better than the baselines, ranging from PMF to DSR.
Compared with other HIN based methods,  HERec adopts a more principled way to leverage HINs for improving recommender systems, which provides better information extraction (a new HIN embedding model) and utilization (an extended MF model).
Moreover, the superiority of the proposed HERec becomes more significant with less training data. In particular, the improvement ratio of HERec over the PMF is up to 40\% with 20\% training data on the Douban Book dataset, which indicates a significant performance boost.
As mentioned previously, the Yelp dataset is very sparse. In this case, even with 60\% training data, the HERec model improves over PMF by about 26\%. As a comparison, with the same training ratio (\ie 60\%), the improvement ratios over PMF are about 29\% in terms of RMSE scores. These results indicate the effectiveness of the proposed approach, especially in a sparse dataset.

(3) Considering the two HERec variants HERec$_{dw}$ and HERec$_{mp}$, it is easy to see HERec$_{dw}$ performs much worse than HERec$_{mp}$.
The major difference lies in the network embedding component. HERec$_{mp}$ adopts the recently proposed HIN embedding method \emph{metapath2vec}~\cite{dong2017metapath2vec}, while HERec$_{dw}$ ignores data heterogeneity
and casts HINs as homogeneous networks for learning. It indicates HIN embedding methods are important to HIN based recommendation.
The major difference between the proposed embedding method and metapath2vec (adopted by HERec$_{mp}$) lies in that we try to transform heterogeneous network information into homogeneous neighborhood, while metapath2vec tries to map all types of entities into the same representation space using heterogeneous sequences.
Indeed, metapath2vec is a general-purpose HIN embedding method, while the proposed embedding method aims to improve the recommendation performance.
Our focus is to learn the embeddings for users and items, while objects of other types are only used as the bridge to construct the homogeneous neighborhood. Based on the results (HERec $>$ HERec$_{mp}$), we argue that it is more effective to perform task-specific HIN embedding for improving the recommendation performance.

%(5) In addition, we can also observe that the superiority of the proposed HERec becomes more significant with less training data.
%For example, on  Douban Movie dataset, using 20\% training data, HERec outperforms PMF up to 18.85\% on RMSE and 18.58\% on MAE.
%It implies that our method is more effective to alleviate the cold-start problem.

%\begin{figure}[t]%[htbp]
%\centering
%\subfigure[RMSE]{
%\begin{minipage}[t]{0.3\linewidth}
%\centering
%\includegraphics[width=4.2cm]{image/cold_start_rmse.pdf}
%\end{minipage}
%}
%\hspace{30pt}
%\subfigure[MAE]{
%\begin{minipage}[t]{0.3\linewidth}
%\centering
%\includegraphics[width=4.2cm]{image/cold_start_mae.pdf}
%\end{minipage}
%}
%\caption{\label{fig_cs}Performance comparison of different methods for cold-start prediction. Improvement ratios over PMF are reported.}
%\end{figure}

%\begin{figure}[t]%[htbp]
%\centering
%\subfigure[RMSE]{
%\begin{minipage}[t]{0.3\linewidth}
%\centering
%\includegraphics[width=4.2cm]{image/reg_rmse.pdf}
%\end{minipage}
%}
%\hspace{40pt}
%\subfigure[MAE]{
%\begin{minipage}[t]{0.3\linewidth}
%\centering
%\includegraphics[width=4.2cm]{image/reg_mae.pdf}
%\end{minipage}
%}
%\caption{\label{fig_reg}Varying parameters $\alpha$ and $\beta$.}
%\end{figure}

\subsection{Detailed Analysis of The Proposed Approach}
In this part, we perform a series of detailed analysis for the proposed approach.

%\begin{figure}[t]%[htbp]
%\centering
%\includegraphics[width=8cm]{image/cold_start_rmse.pdf}
%\caption{\label{fig_cs}Performance comparison of different methods for cold-start prediction on Douban Movie dataset. $y$-axis denotes the improvement ratio over PMF.}
%\end{figure}

\begin{figure*}
\centering
\subfigure[Douban Movie]{
\begin{minipage}[b]{0.3\textwidth}
\includegraphics[width=1\textwidth]{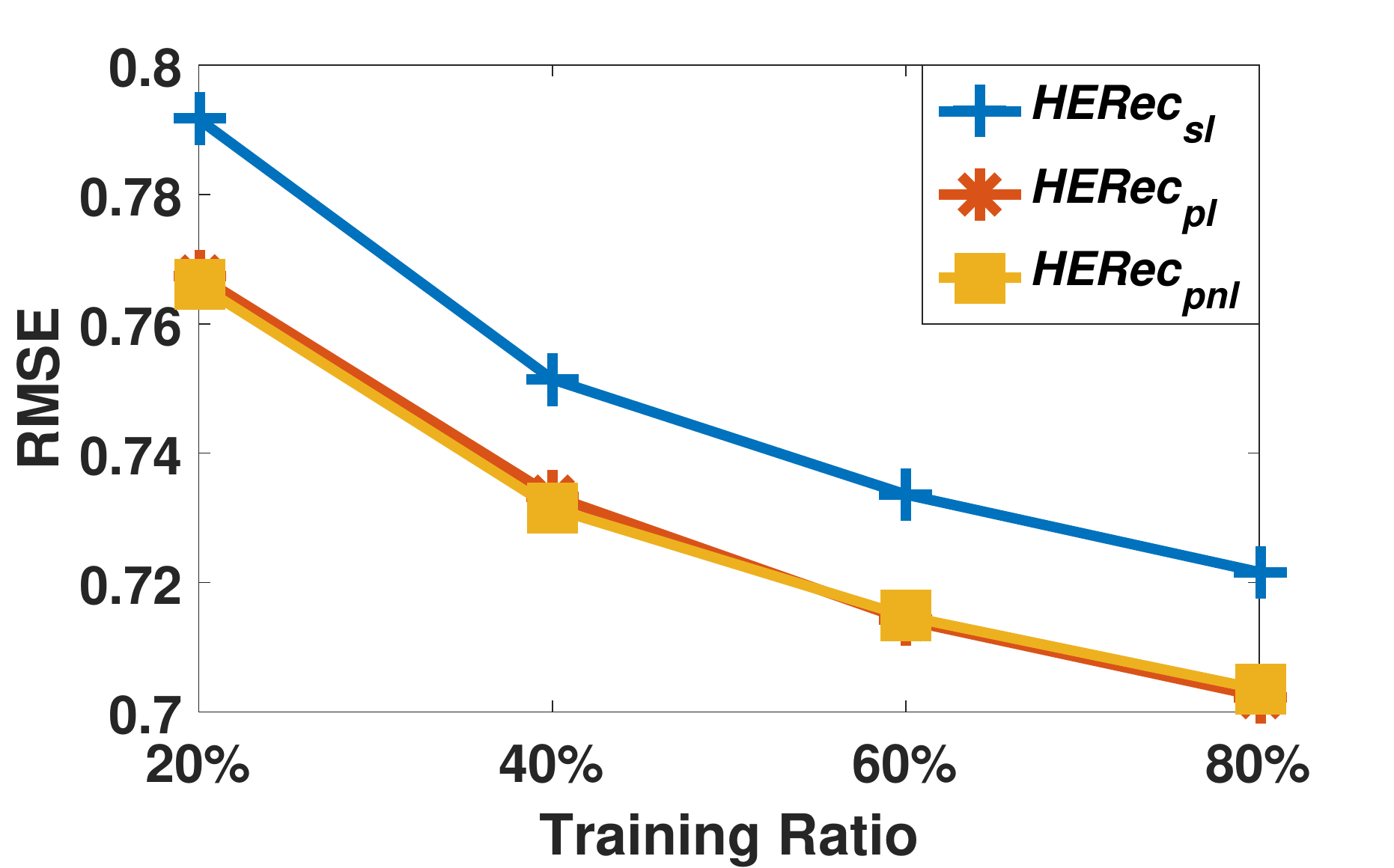} \\
\includegraphics[width=1\textwidth]{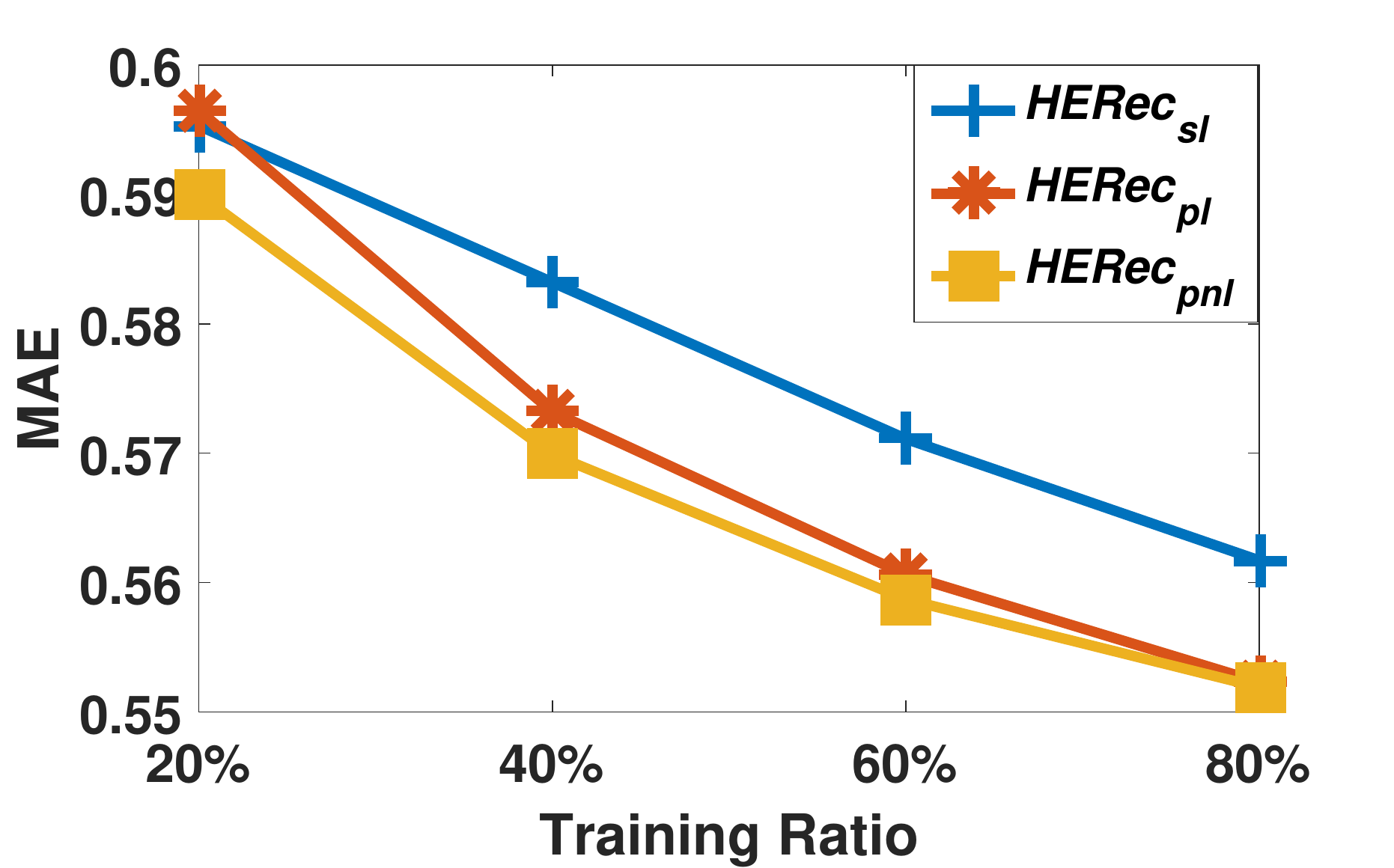}
\end{minipage}
}
\subfigure[Douban Book]{
\begin{minipage}[b]{0.3\textwidth}
\includegraphics[width=1\textwidth]{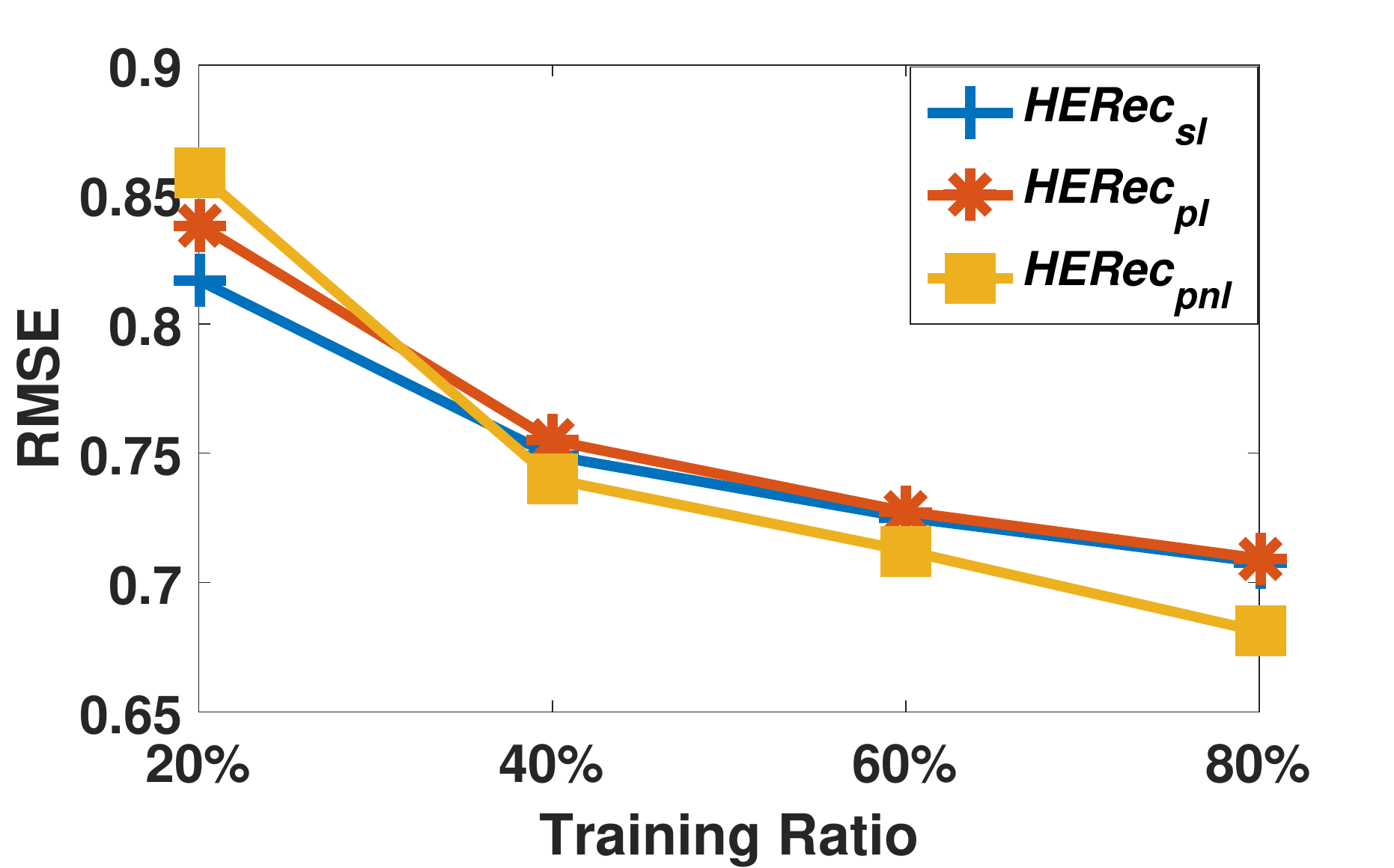} \\
\includegraphics[width=1\textwidth]{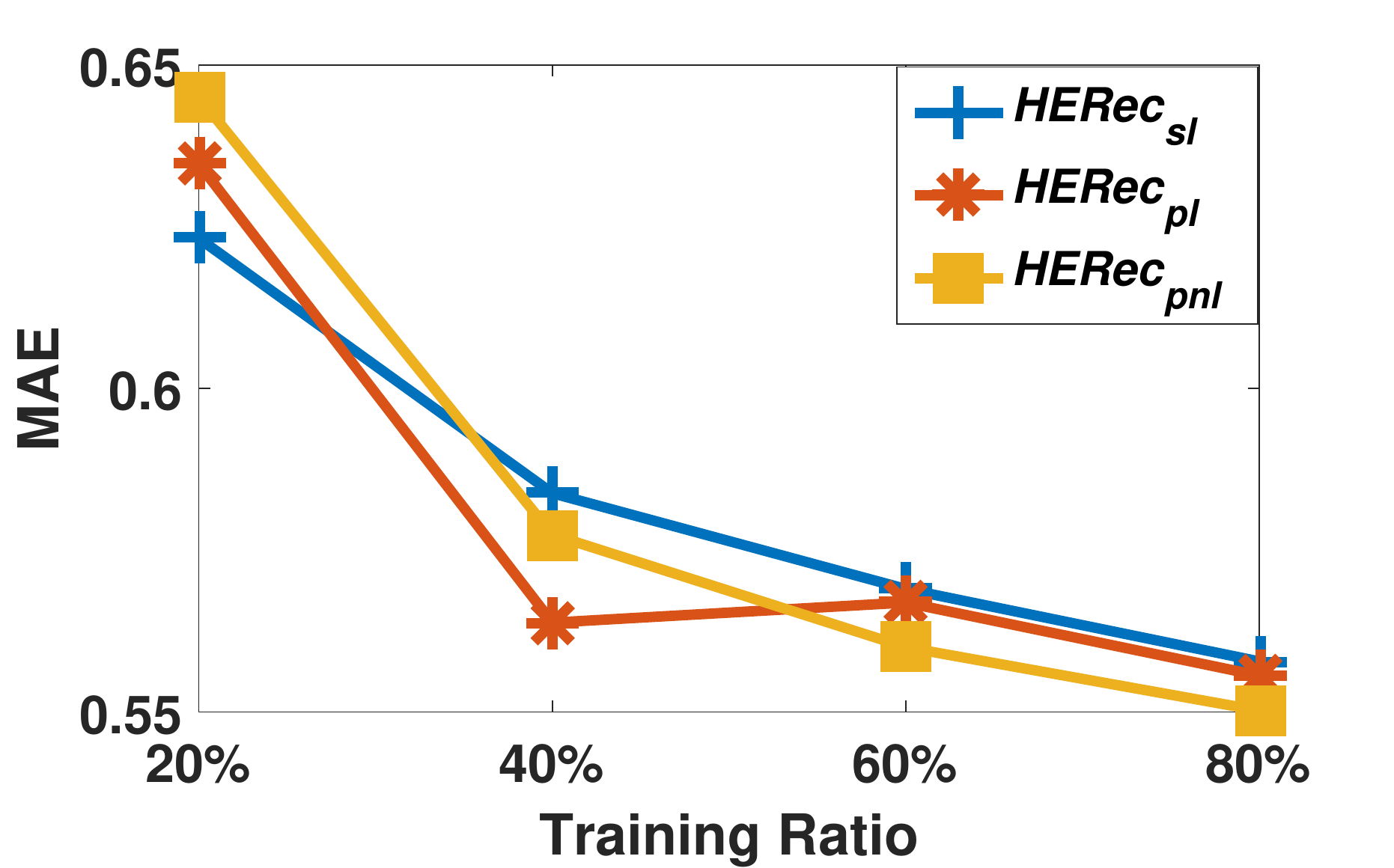}
\end{minipage}
}
\subfigure[Yelp]{
\begin{minipage}[b]{0.3\textwidth}
\includegraphics[width=1\textwidth]{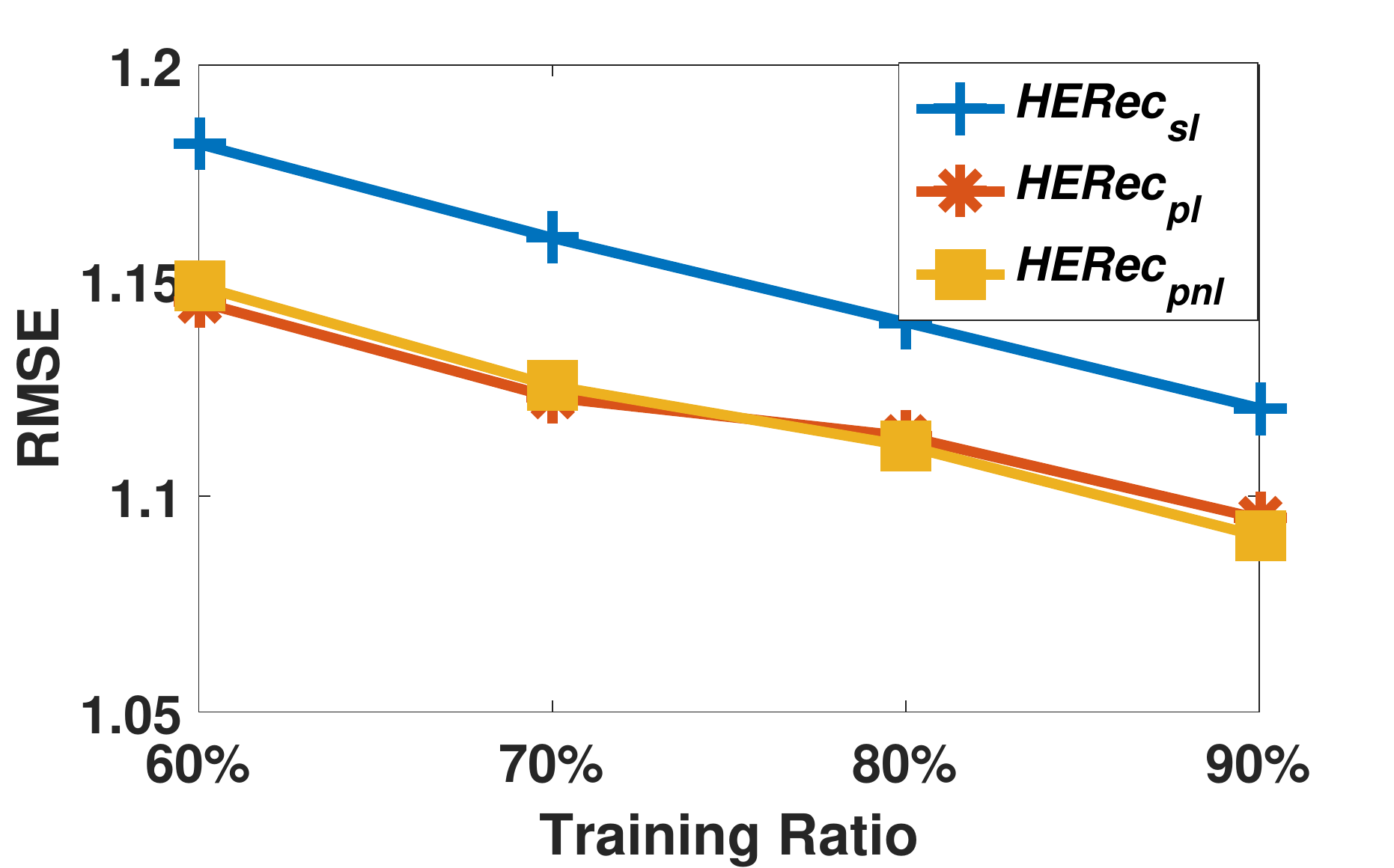} \\
\includegraphics[width=1\textwidth]{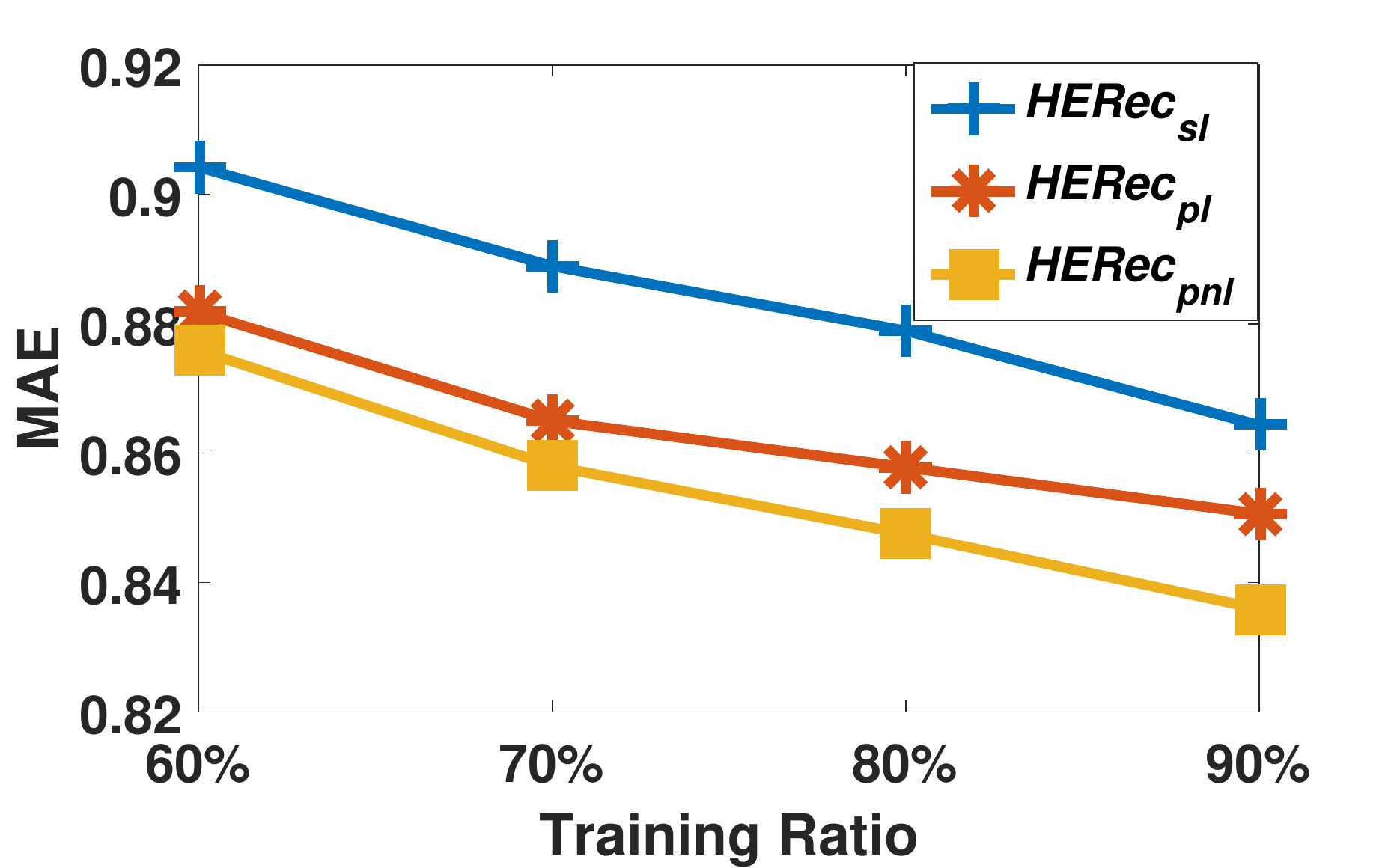}
\end{minipage}
}
\caption{\label{fig_fusion}Performance comparison of different fusion functions on three datasets.}
\end{figure*}

\begin{figure*}
\centering
\subfigure[Douban Movie]{
\begin{minipage}[b]{0.3\textwidth}
\includegraphics[width=1\textwidth]{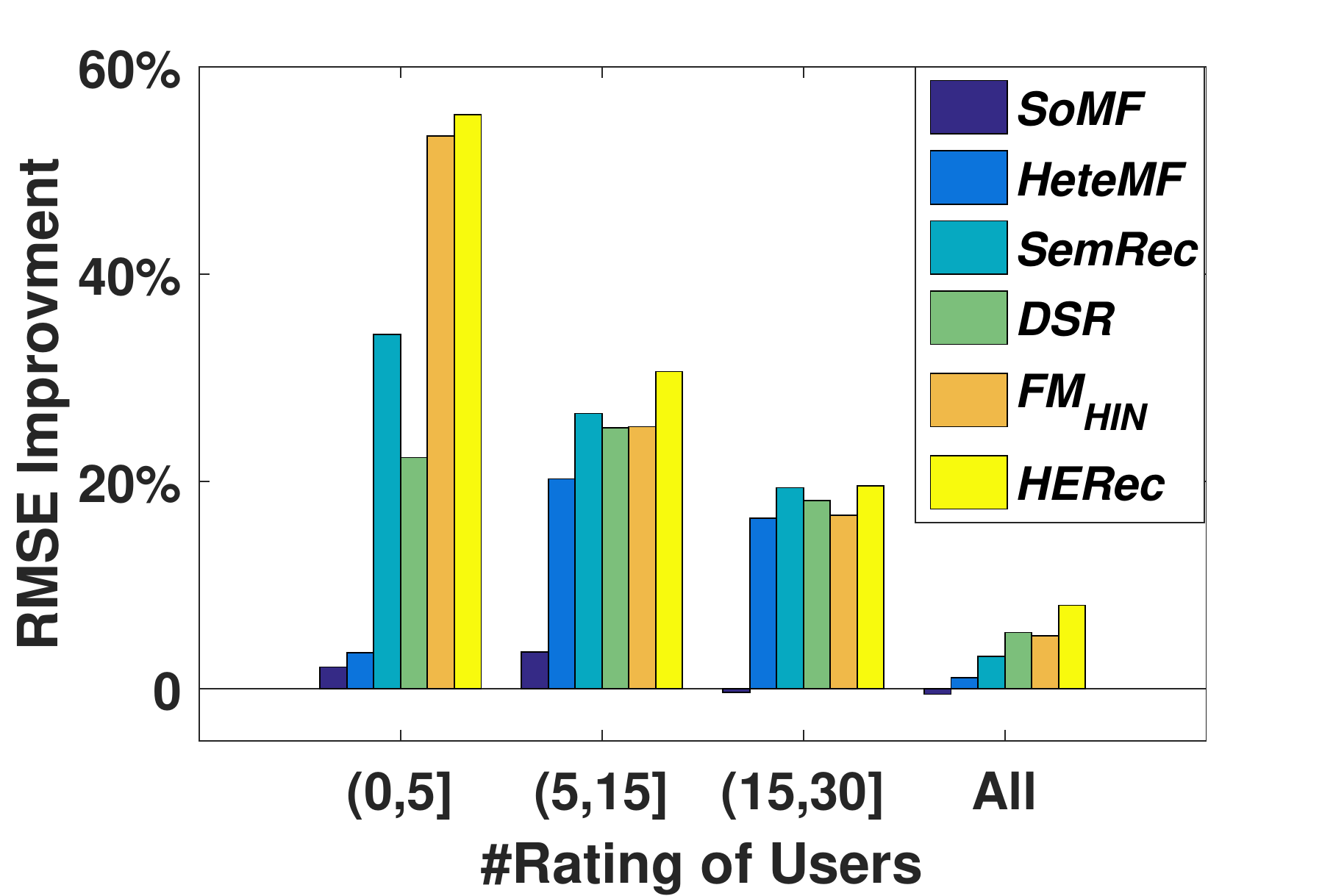} \\
\includegraphics[width=1\textwidth]{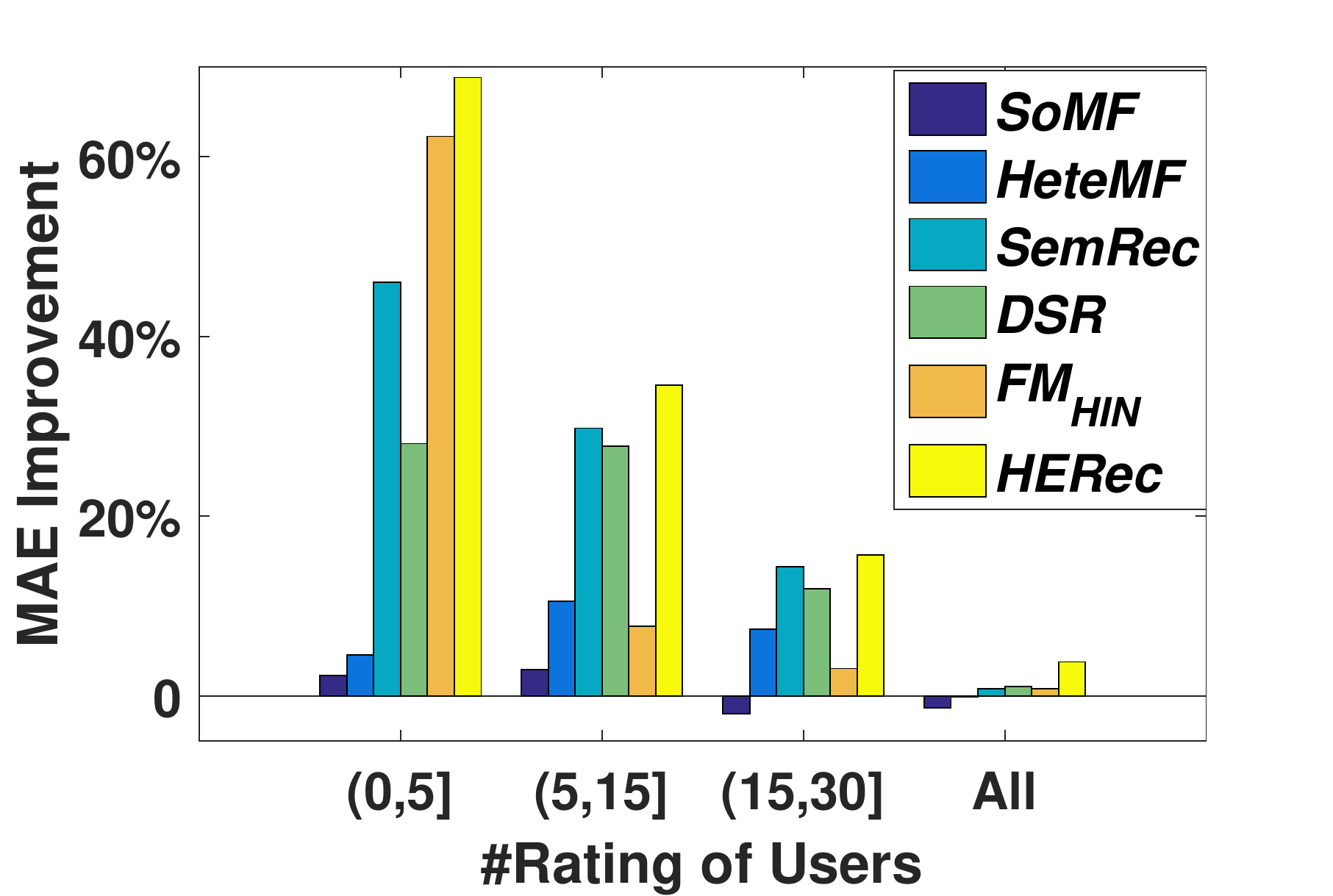}
\end{minipage}
}
\subfigure[Douban Book]{
\begin{minipage}[b]{0.3\textwidth}
\includegraphics[width=1\textwidth]{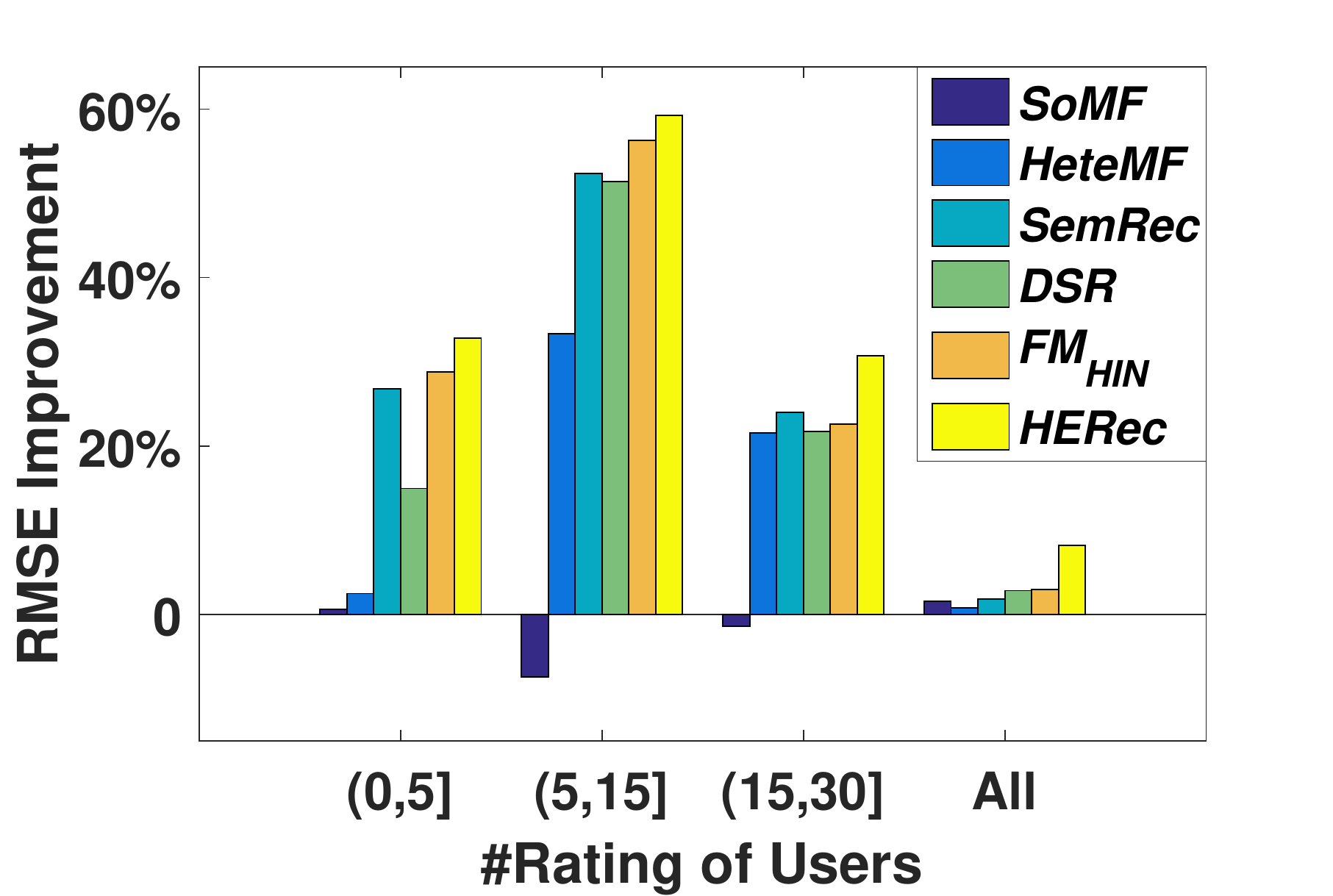} \\
\includegraphics[width=1\textwidth]{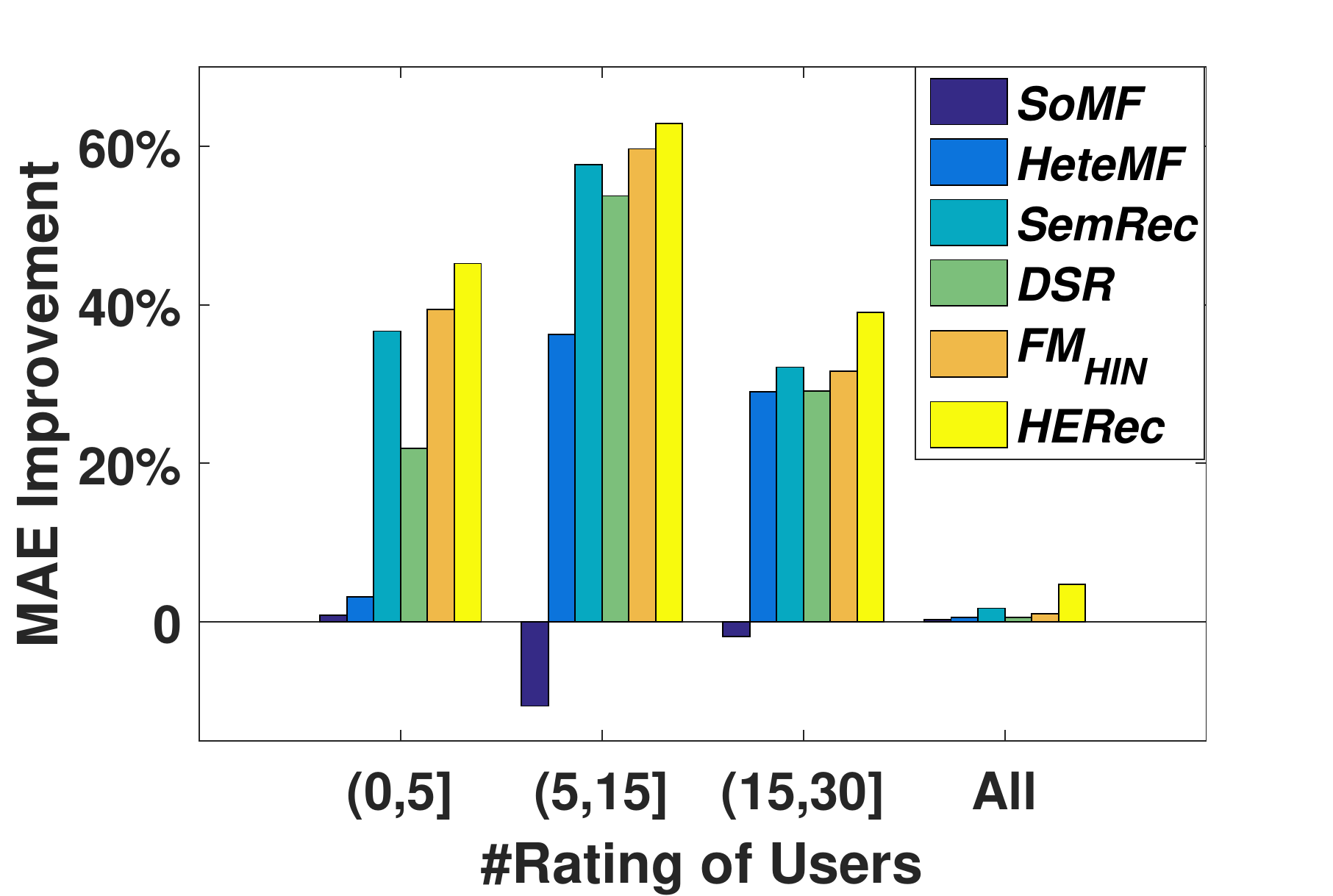}
\end{minipage}
}
\subfigure[Yelp]{
\begin{minipage}[b]{0.3\textwidth}
\includegraphics[width=1\textwidth]{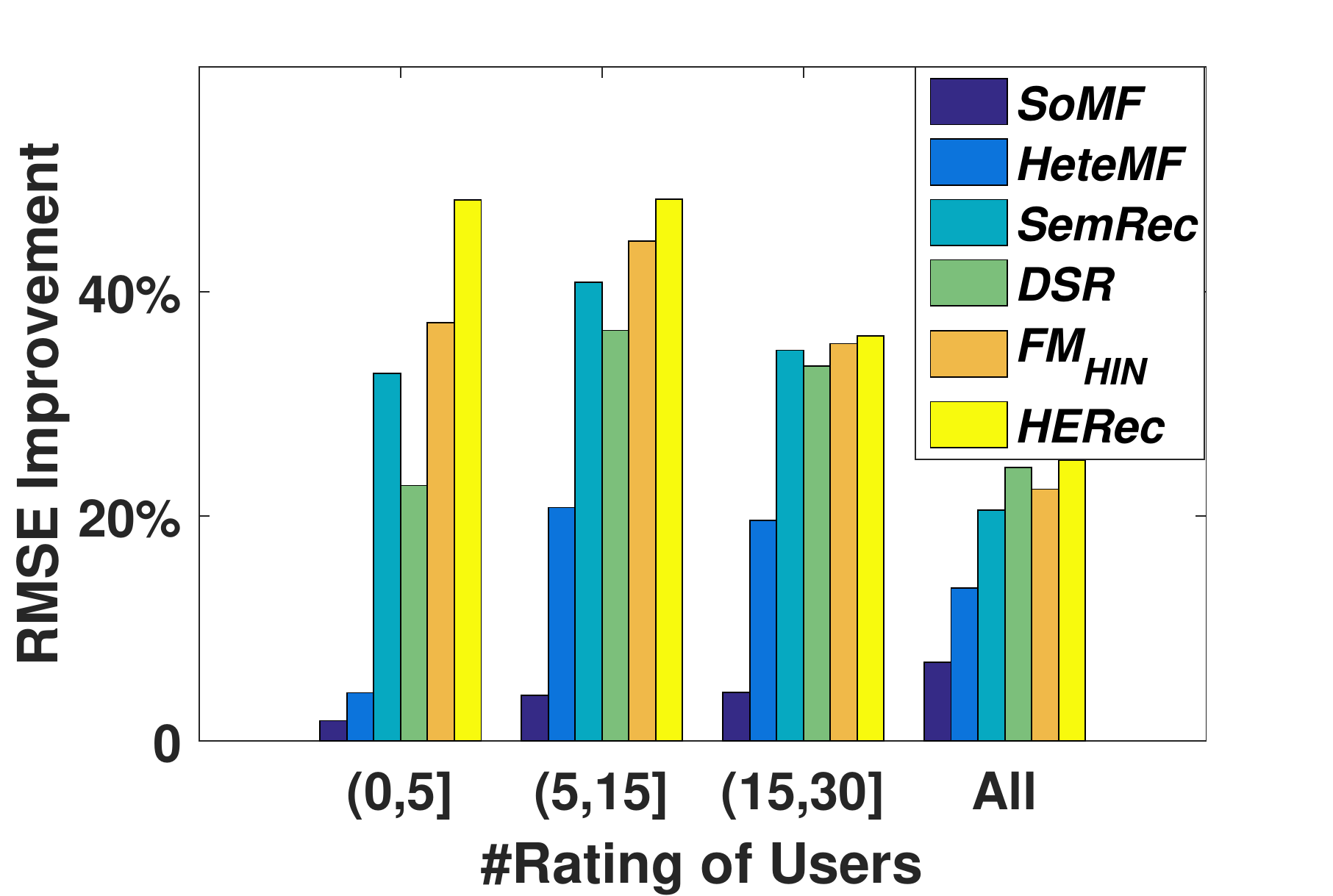} \\
\includegraphics[width=1\textwidth]{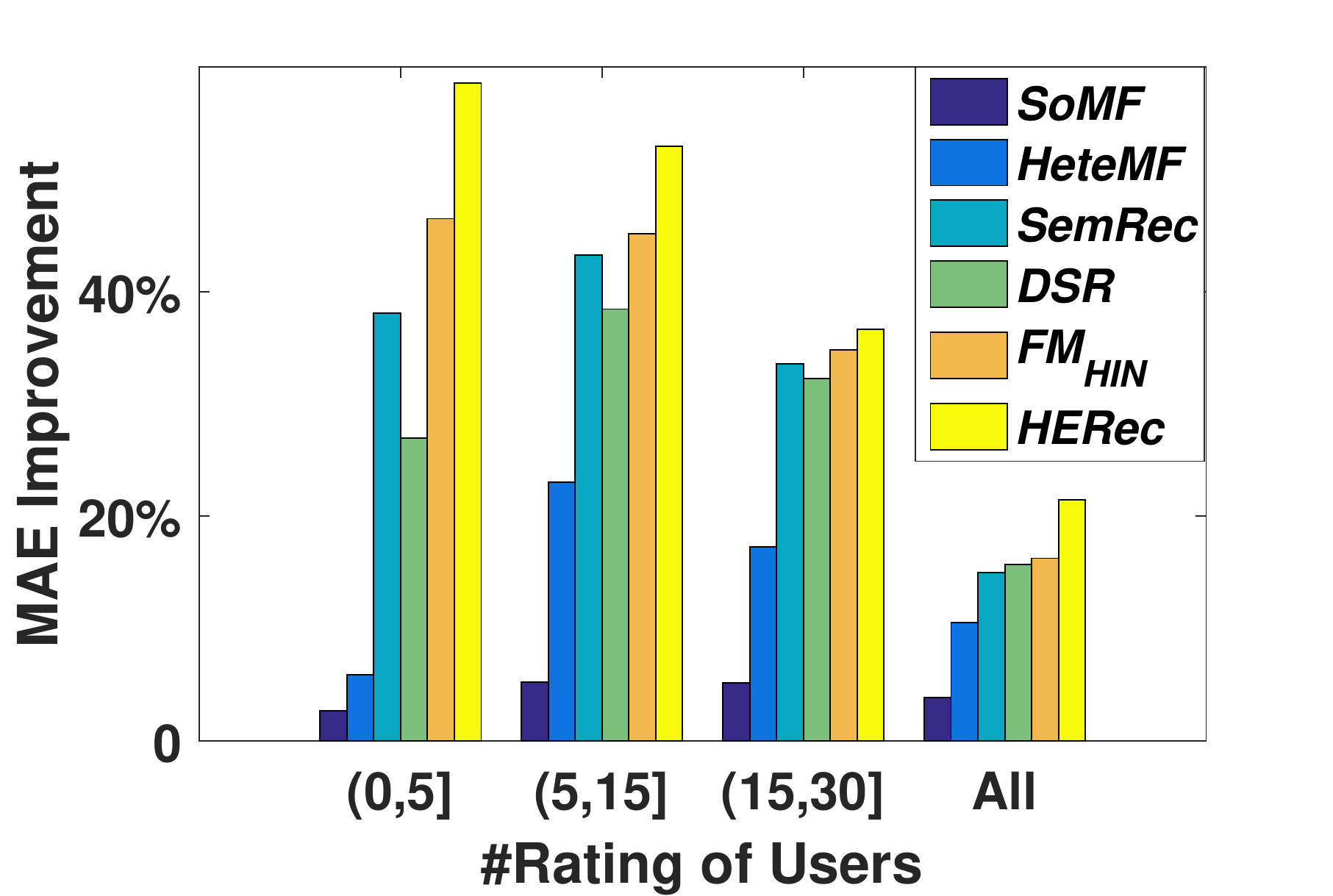}
\end{minipage}
}
\caption{\label{fig_cs}Performance comparison of different methods for cold-start prediction on three datasets. $y$-axis denotes the improvement ratio over PMF.}
\end{figure*}

\subsubsection{Selection of Different Fusion Functions}
HERec requires a principled fusion way to transform node embeddings into a more suitable form that is useful to enhance recommendation performance. Therefore, we will discuss the impact of different fusion functions on the recommendation performance.
For convenience, we call the HERec variant with the simple linear fusion function (Eq.~\ref{eq-slf}) as HERec$_{sl}$, the variant with personalized linear fusion function  (Eq.~\ref{eq-plf}) as HERec$_{pl}$,  and the variant with personalized non-linear fusion function (Eq.~\ref{eq-pnlf}) as HERec$_{pnl}$.
We present the performance comparison of the three variants of HERec on the three datasets in Fig.~\ref{fig_fusion}.

As shown in Fig.~\ref{fig_fusion}, we can find that overall performance ranking is as follows: HERec$_{pnl}$ $>$ HERec$_{pl}$ $>$ HERec$_{sl}$. Among the three variants, the simple linear fusion function performs the worst, as it ignores the personalization and non-linearity. Indeed, users are likely to have varying preferences over meta-paths~\cite{shi2015semantic}, which should be considered in meta-path based methods. The personalization factor improves the performance significantly. As a comparison, the performance improvement of
HERec$_{pnl}$ over  HERec$_{pl}$ is relatively small. A possible reason is that the incorporation of personalized combination parameters increases the capability of linear models. Nonetheless, HERec$_{pnl}$ still performs the best by considering personalization and non-linear transformation.
In a complicated recommender setting, HIN embeddings may not be directly applicable in recommender systems, where a non-linear mapping function is preferred.

Since HERec$_{pnl}$ is the best variant of the proposed model, in what follows, HERec will use the personalized non-linear fusion function, \ie HERec$_{pnl}$ by default.

\subsubsection{Cold-start Prediction}
HINs are particularly useful to improve cold-start prediction, where there are fewer rating records but heterogeneous context information is available. We consider studying the performance \emph{w.r.t.} different cold-start degrees, \ie the rating sparsity.
To test it, we first  categorize ``cold" users into three groups according to the numbers of their rating records, \ie $(0, 5]$, $(5, 15]$ and $(15, 30]$. It is easy to see that the case for the first group is the most difficult, since users from this group have fewest rating records.
Here, we only select the baselines which use HIN based information for recommendation, including SoMF, HeteMF, SemRec, DSR and FM$_{HIN}$.
We present the performance comparison of different methods in Fig.~\ref{fig_cs}. For convenience, we report the improvement ratios \emph{w.r.t.} PMF.
Overall, all the comparison methods are better than PMF (\ie a positive $y$-axis value). The proposed method performs the best among all the methods, and the improvement over PMF becomes more significant for users with fewer rating records. The results indicate that HIN based information is effective to improve the recommendation performance, and the proposed HERec method can effectively utilize HIN information in a more principled way.

%\begin{figure}[htbp]
%%\centering
%\subfigure[Yelp]{
%\begin{minipage}[t]{0.3\linewidth}
%\centering
%\includegraphics[width=4.2cm]{image/metapath_yelp.pdf}
%\end{minipage}
%}
%\hspace{40pt}
%\subfigure[Douban Movie]{
%\begin{minipage}[t]{0.3\linewidth}
%\centering
%\includegraphics[width=4.2cm]{image/metapath_douban.pdf}
%\end{minipage}
%}
%\caption{\label{fig_metapath}Performance change of HERec when gradually incorporating meta-paths.}
%\end{figure}

%\begin{figure}[htbp]
%%\centering
%\subfigure[Yelp]{
%\begin{minipage}[t]{0.5\textwidth}
%\centering
%\includegraphics[width=8cm]{image/metapath_yelp.pdf}
%\end{minipage}
%}
%\subfigure[Douban Movie]{
%\begin{minipage}[t]{0.5\textwidth}
%\centering
%\includegraphics[width=8cm]{image/metapath_douban.pdf}
%\end{minipage}
%}
%\caption{\label{fig_metapath}Performance change of HERec when gradually incorporating meta-paths.}
%\end{figure}

\begin{figure*}[htbp]
\centering
\subfigure[Douban Movie]{
\begin{minipage}[t]{0.3\textwidth}
\centering
\includegraphics[width=1\textwidth]{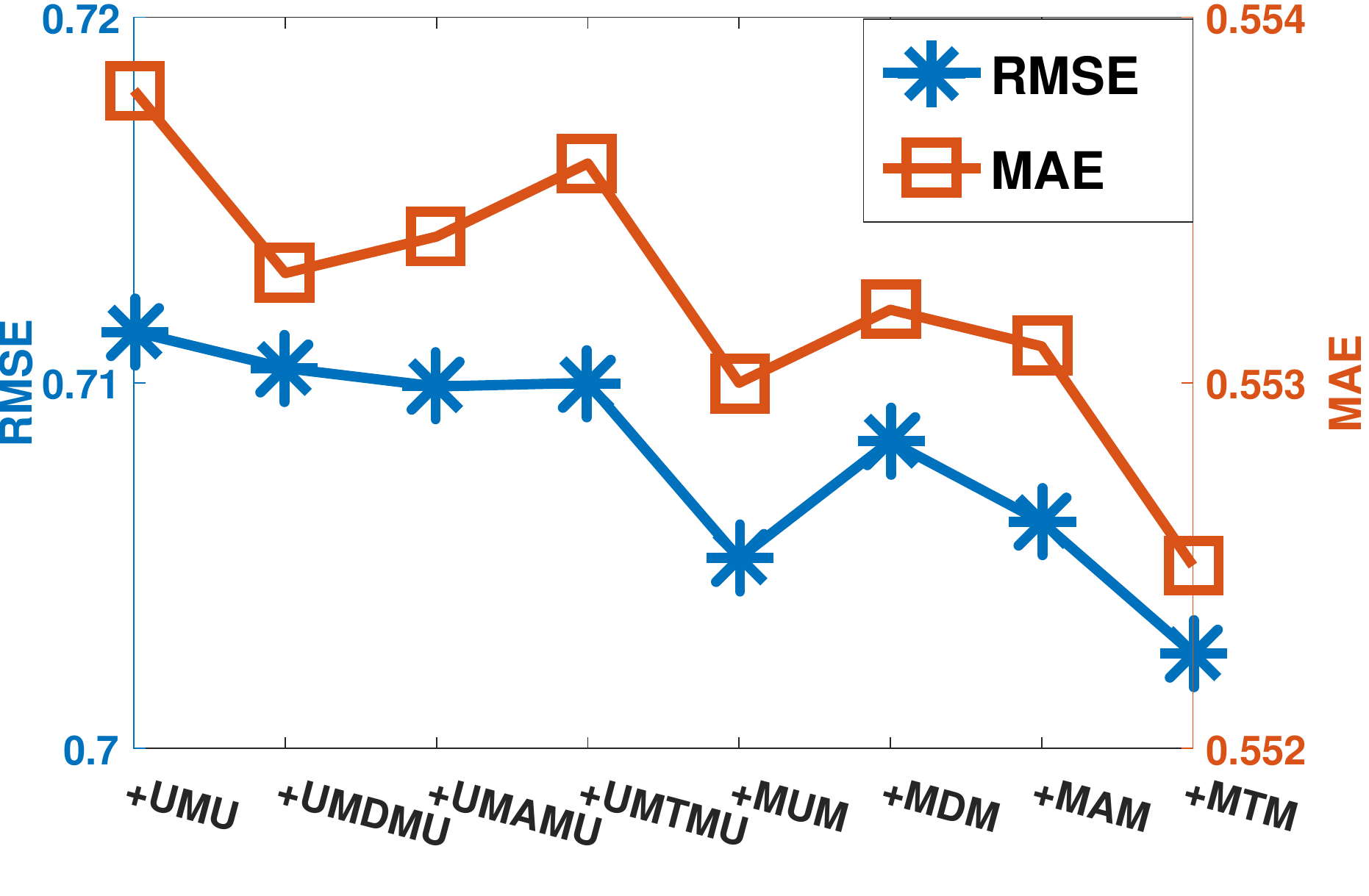}
\end{minipage}
}
\subfigure[Douban Book]{
\begin{minipage}[t]{0.3\textwidth}
\centering
\includegraphics[width=1.05\textwidth]{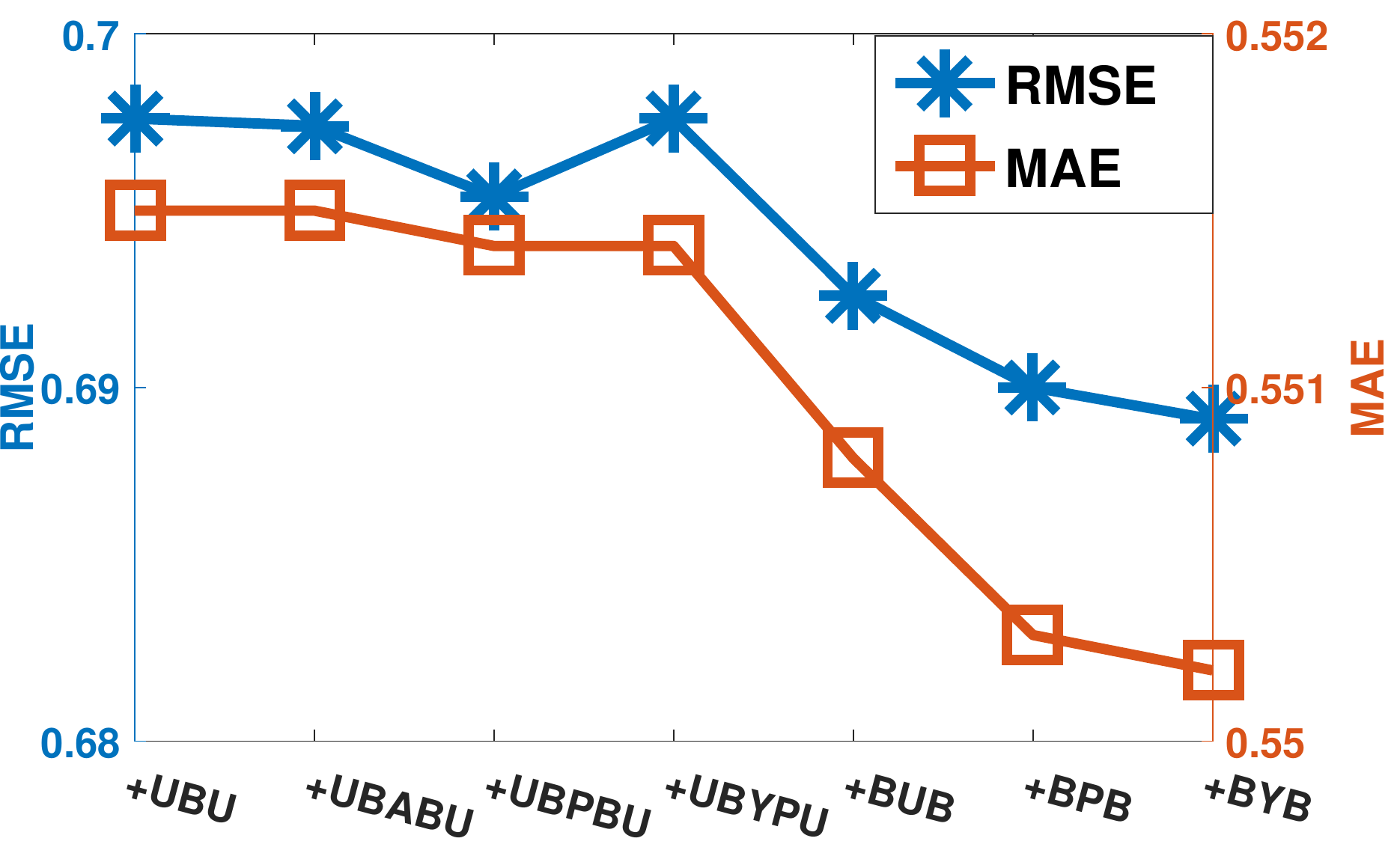}
\end{minipage}
}
\subfigure[Yelp]{
\begin{minipage}[t]{0.3\textwidth}
\centering
\includegraphics[width=1\textwidth]{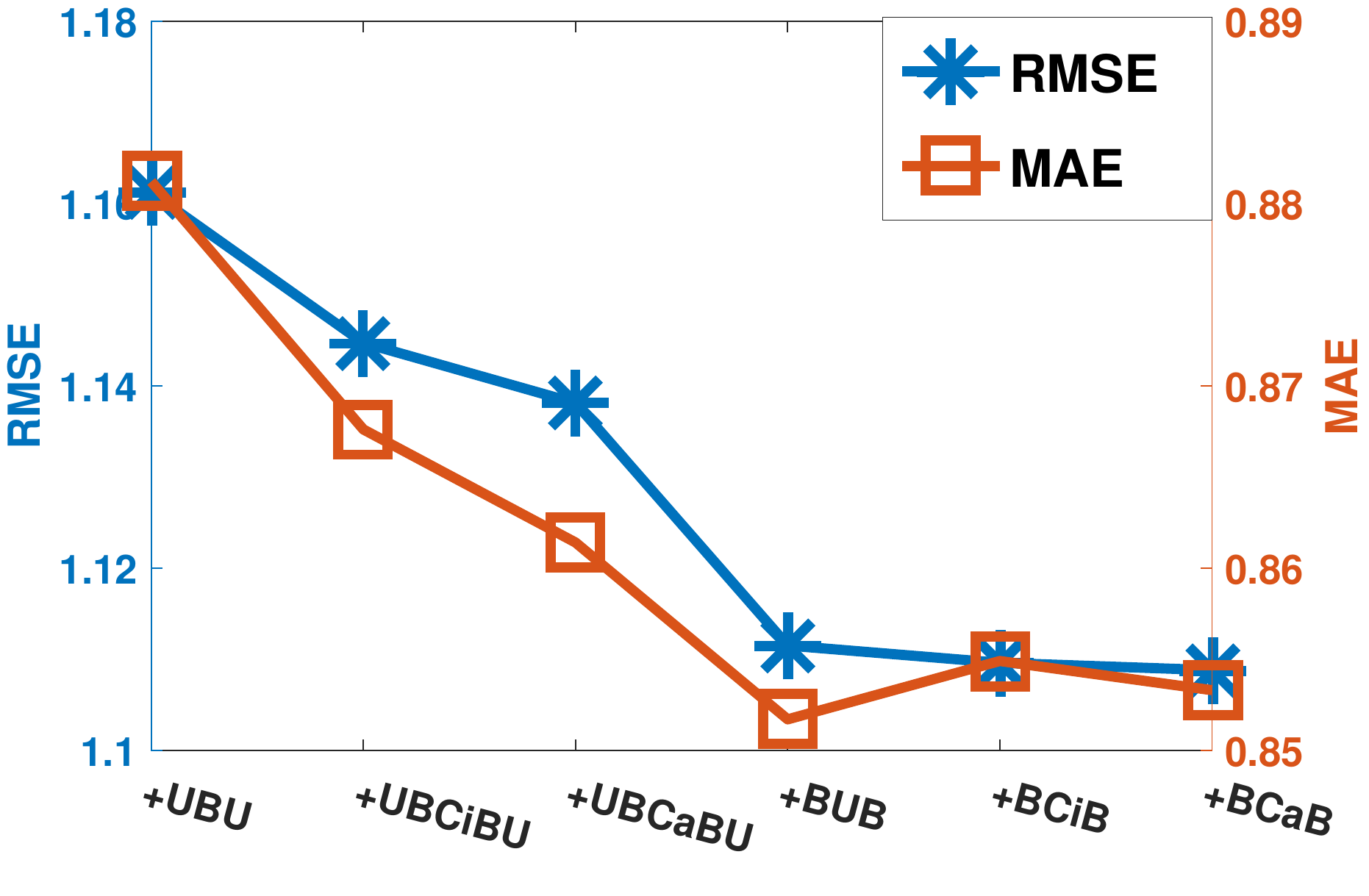}
\end{minipage}
}
\caption{\label{fig_metapath}Performance change of HERec when gradually incorporating meta-paths.}
\end{figure*}

\subsubsection{Impact of Different Meta-Paths}
%Our approach depends on the selection of useful meta-paths.
As shown in Table~\ref{tab_metapath}, the proposed approach uses a selected set of meta-paths.
To further analyze the impact of different meta-paths, we gradually incorporate these meta-paths into the proposed approach and check the performance change.
In Fig. \ref{fig_metapath}, we can observe that generally the performance improves (\ie becoming smaller) with the incorporation of more meta-paths.
Both meta-paths starting with the user type and item type are useful to improve the performance.
However, it does not always yield the improvement with more meta-paths, and the performance slightly fluctuates.
The reason is that some meta-paths may contain noisy  or conflict information with existing ones.
Another useful observation is that the performance quickly achieves a relatively good performance with the incorporation of only a few meta-paths.
This confirms previous finding~\cite{shi2015semantic}:  a small number of high-quality meta-paths are able to lead to large performance improvement. Hence, as mentioned before, we can effectively control the model complexity by selecting just a few meta-paths.

\begin{figure*}[t]%[htbp]
\centering
\subfigure[Douban Movie]{
\begin{minipage}[t]{0.3\textwidth}
\centering
\includegraphics[width=1\textwidth]{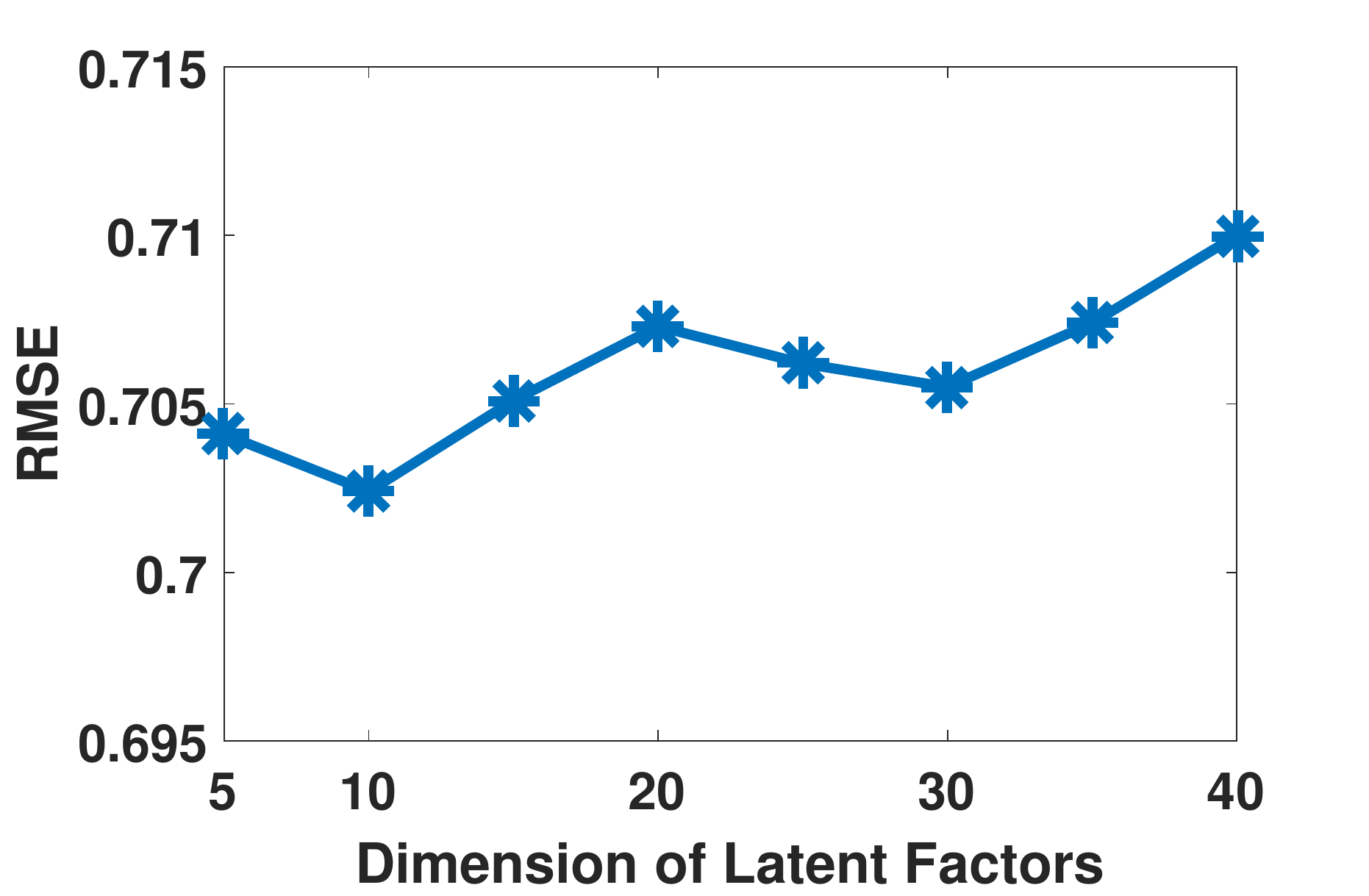}
\end{minipage}
}
\subfigure[Douban Book]{
\begin{minipage}[t]{0.3\textwidth}
\centering
\includegraphics[width=1\textwidth]{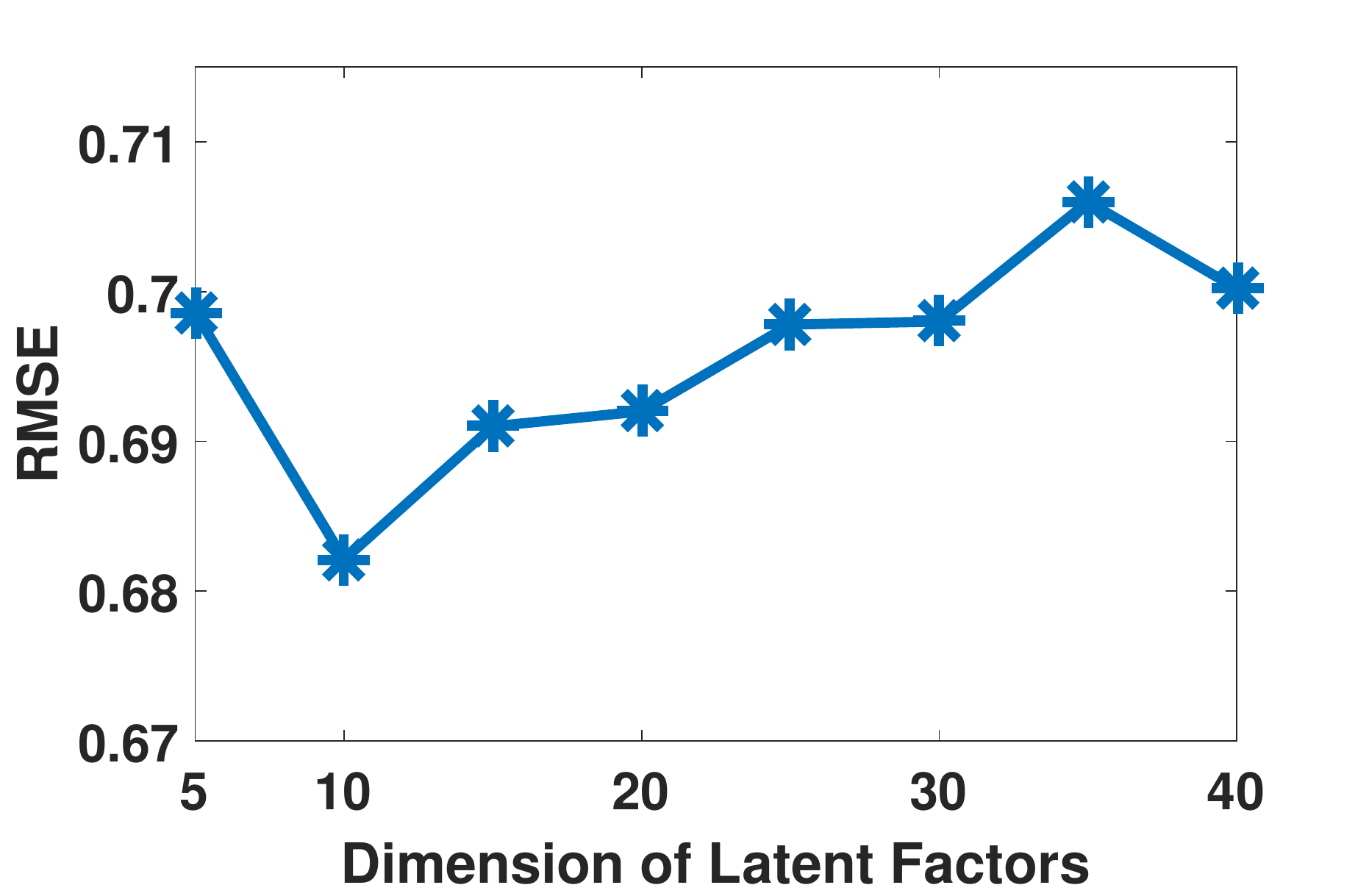}
\end{minipage}
}
\subfigure[Yelp]{
\begin{minipage}[t]{0.3\textwidth}
\centering
\includegraphics[width=1\textwidth]{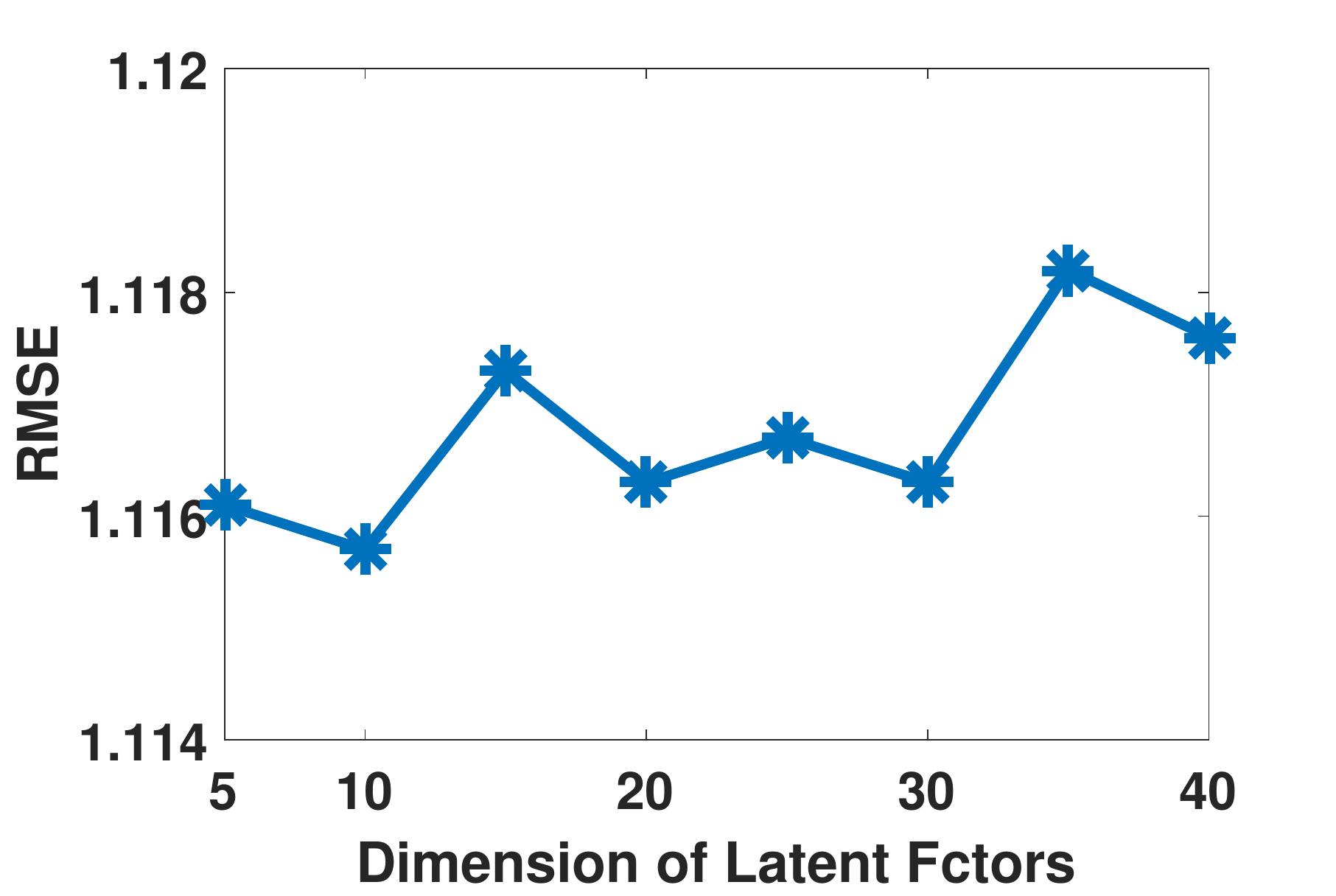}
\end{minipage}
}
\caption{\label{fig_factor}Performance with respect to the dimension of latent factors on three datasets.}
\end{figure*}

\begin{figure*}[t]%[htbp]
\centering
\subfigure[Douban Movie]{
\begin{minipage}[t]{0.3\textwidth}
\centering
\includegraphics[width=1\textwidth]{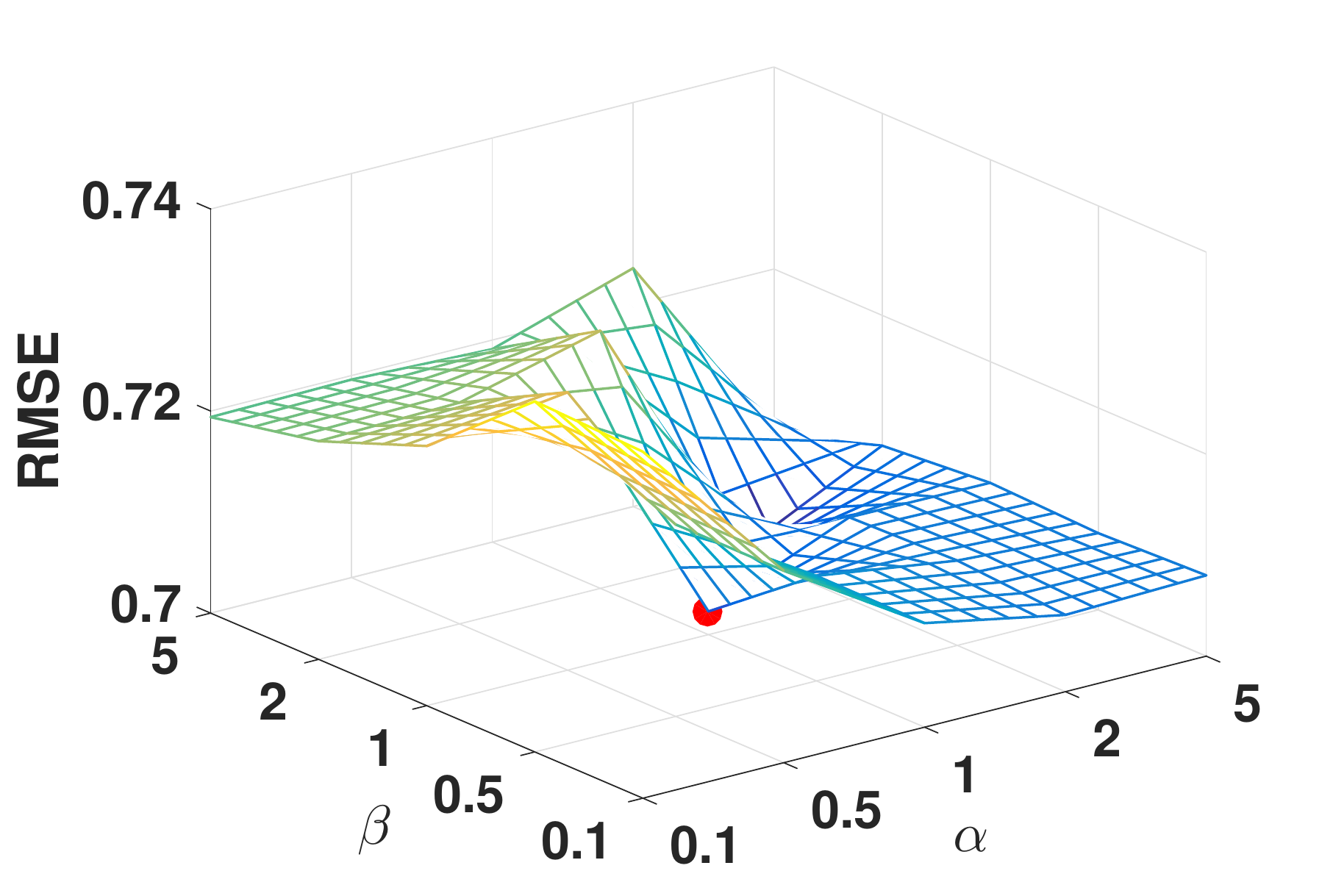}
\end{minipage}
}
\subfigure[Douban Book]{
\begin{minipage}[t]{0.3\textwidth}
\centering
\includegraphics[width=1\textwidth]{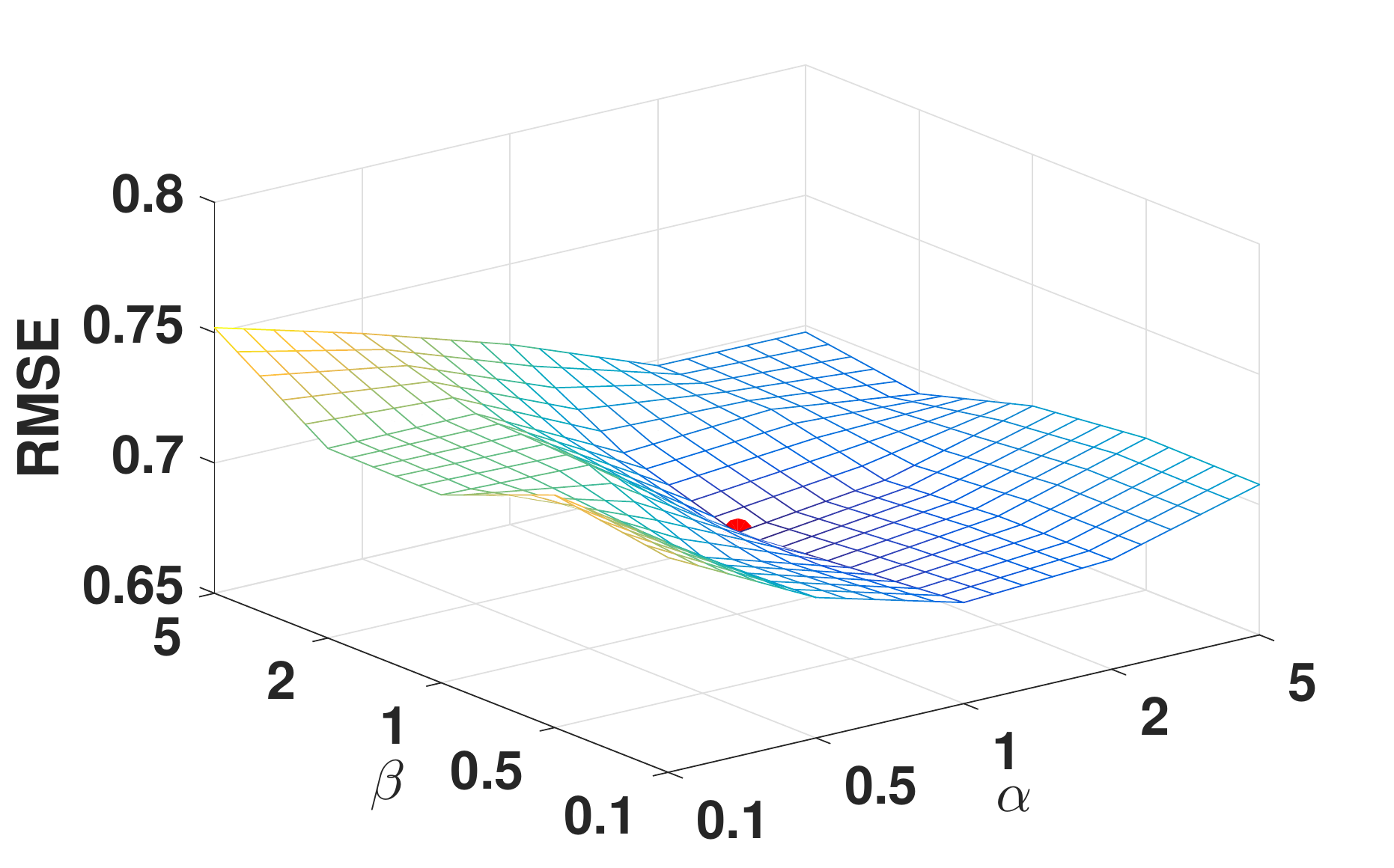}
\end{minipage}
}
\subfigure[Yelp]{
\begin{minipage}[t]{0.3\textwidth}
\centering
\includegraphics[width=1\textwidth]{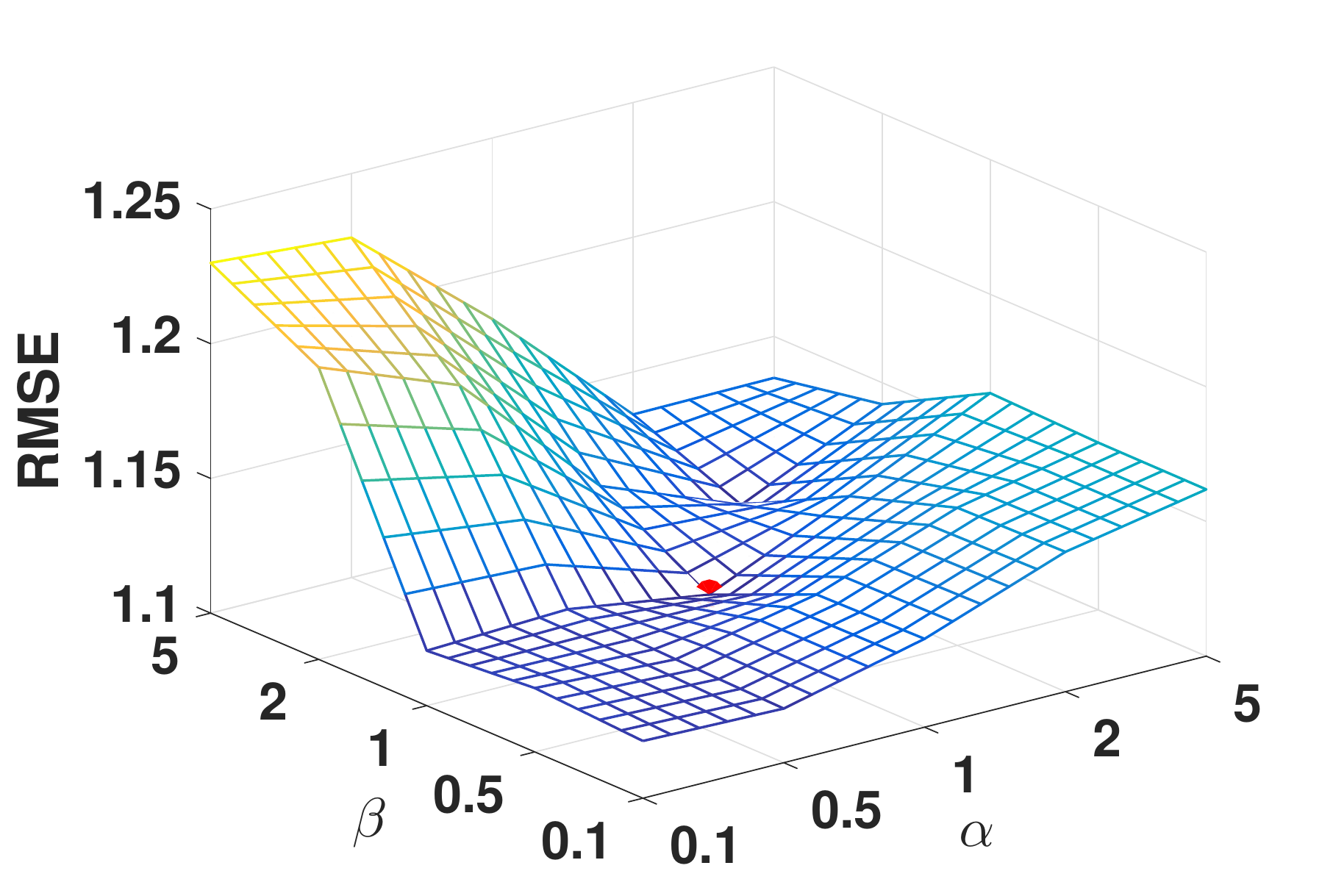}
\end{minipage}
}
\caption{\label{fig_reg}Varying parameters $\alpha$ and $\beta$ on the three datasets.}
\end{figure*}

\begin{figure*}[t]%[htbp]
\centering
\subfigure[Douban Movie]{
\begin{minipage}[t]{0.3\textwidth}
\centering
\includegraphics[width=1.06\textwidth]{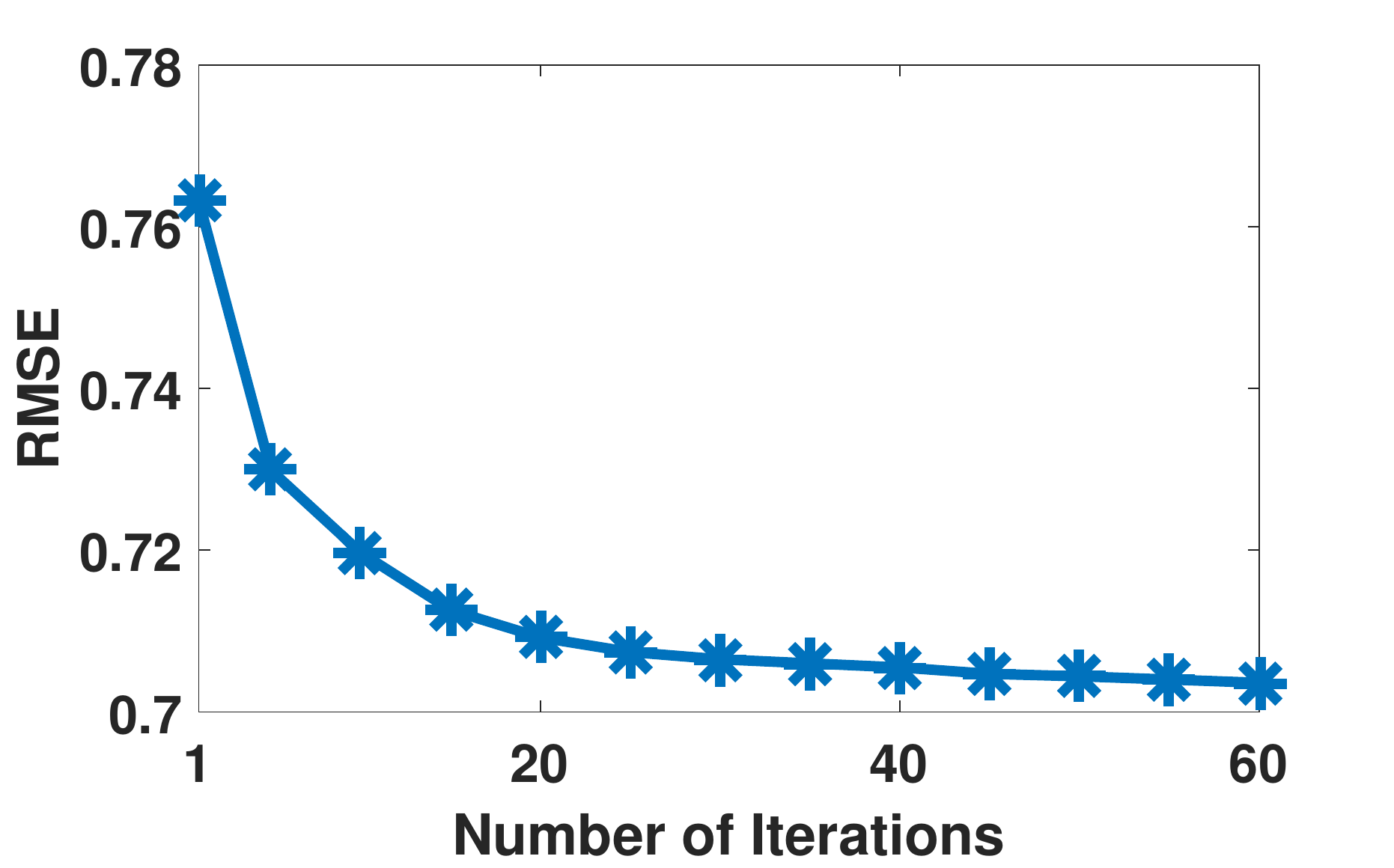}
\end{minipage}
}
\subfigure[Douban Book]{
\begin{minipage}[t]{0.3\textwidth}
\centering
\includegraphics[width=1\textwidth]{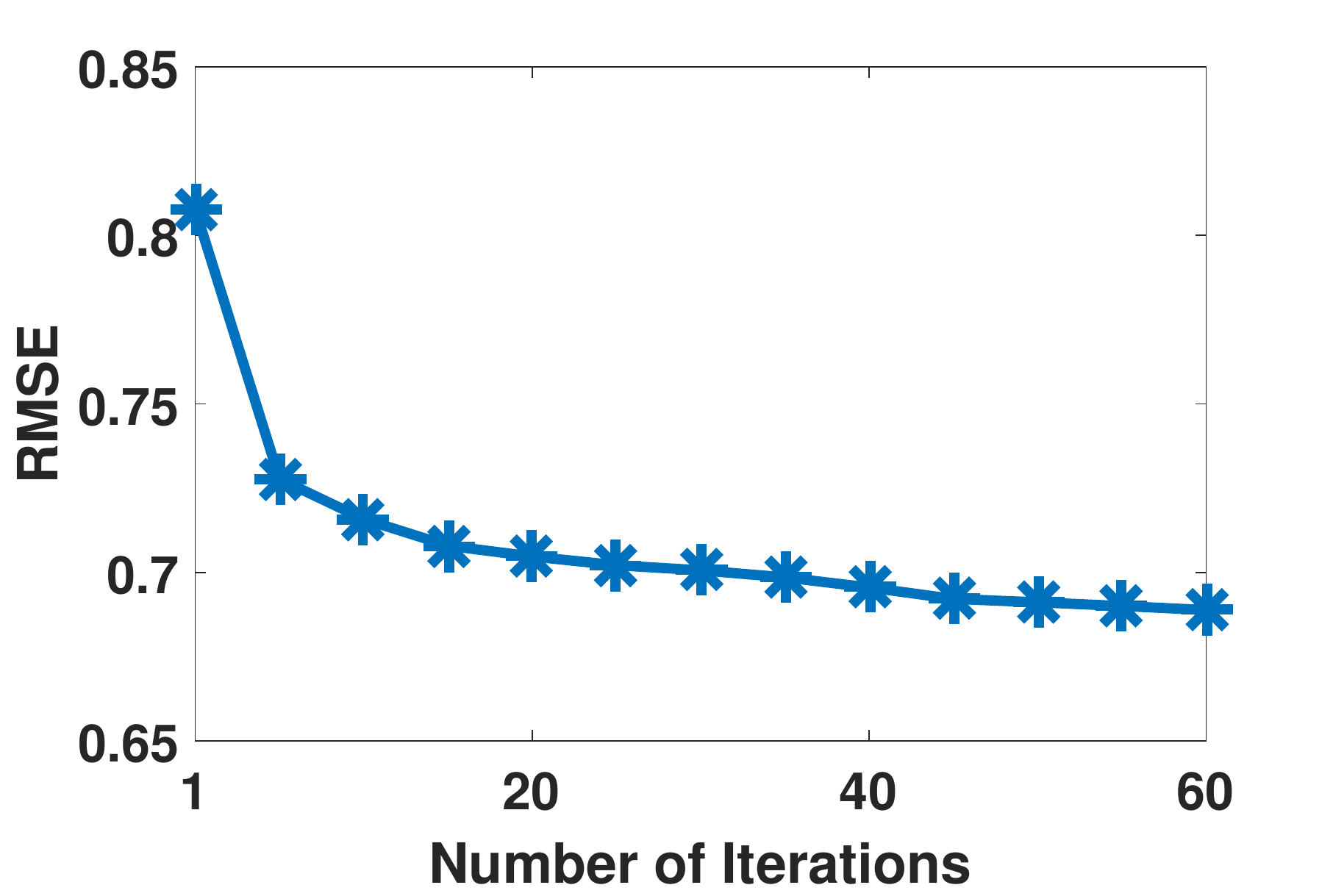}
\end{minipage}
}
\subfigure[Yelp]{
\begin{minipage}[t]{0.3\textwidth}
\centering
\includegraphics[width=1\textwidth]{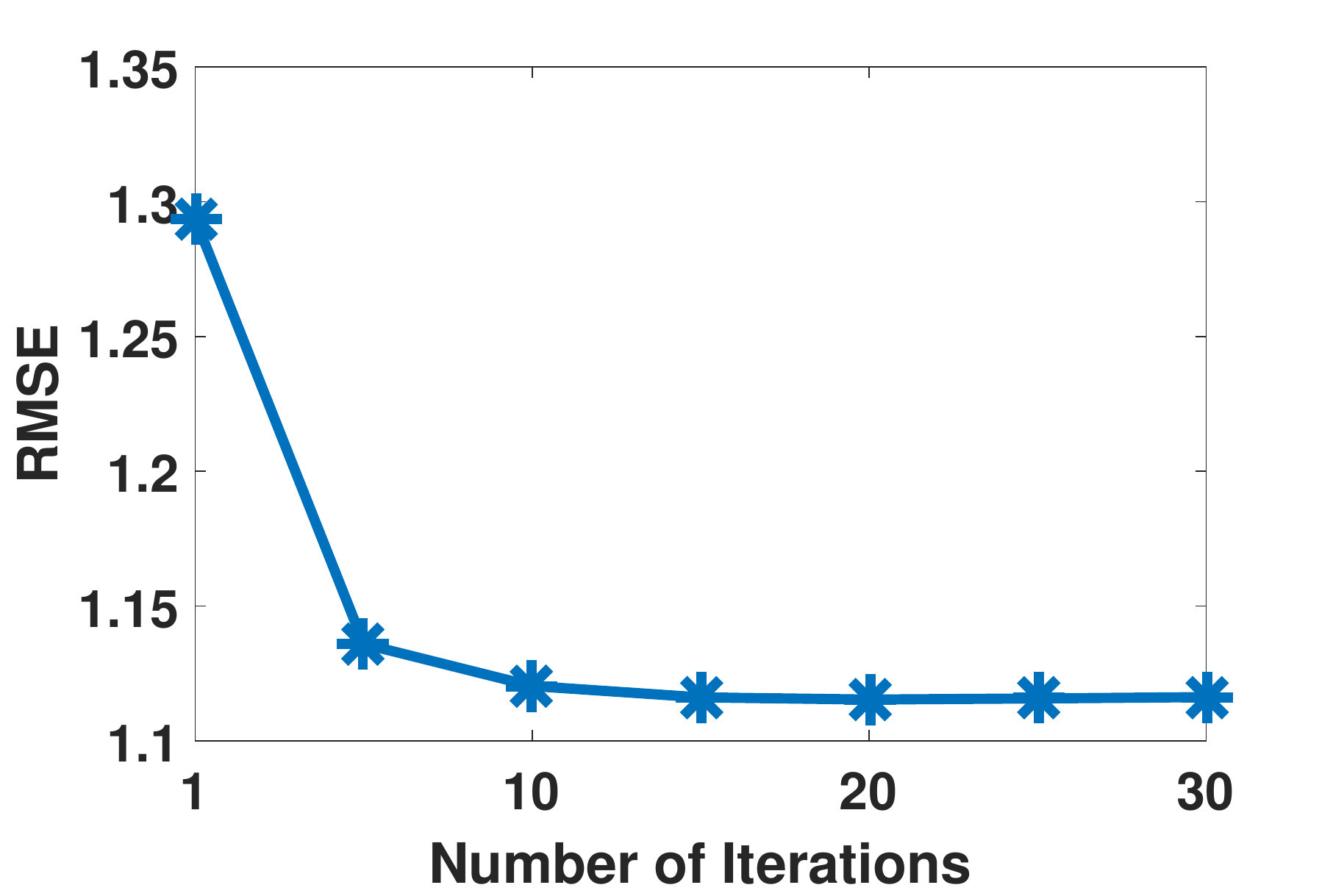}
\end{minipage}
}
\caption{\label{fig_iter}Performance with respect to the number of iterations on three datasets.}
\end{figure*}

\subsubsection{Parameter Tuning}
For matrix factorization based methods, an important parameter to tune is the number of latent factors. The proposed model also involves such a parameter.
We vary it from 5 to 40 with a step of 5, and examine how the performance changes \emph{w.r.t} the number of latent factors.
We present the tuning results in Fig.~\ref{fig_factor}. As we can see, using 10 latent factors yields the best performance, indicating that
the number of latent factors should be set to a small number.

Next, we fix the number of latent factors as ten, and tune two other parameters $\alpha$ and $\beta$ (Eq.~\ref{eq-predictor}), which are used to integrate different terms as weights.
Now we examine how they influence the model performance. For both parameters, we vary them in the set of $\{0.1, 0.5, 1, 2\}$.
 As shown in Fig. \ref{fig_reg}, the optimal performance is obtained near $(1,1)$, \ie both $\alpha$ and $\beta$ are around 1. The results show that
 the HIN embeddings from both the user and item sides are important to improve the prediction performance.
 Overall, the change trend is smooth, indicating that the proposed model is not very sensitive to the two parameters.

 Finally, we study the performance change \emph{w.r.t.} the number of iterations. As shown in Fig.~\ref{fig_iter},
 we can see that the proposed model has a fast convergence rate, and about 40 to 60 iterations are required for dense datasets (\ie Douban Movie and Book), while about 20 iterations
 are required for sparse datasets (\ie Yelp).
%
%\begin{figure}[htbp]
%%\centering
%\subfigure[RMSE]{
%\begin{minipage}[t]{0.5\textwidth}
%\centering
%\includegraphics[width=8cm]{image/window_size_mae.pdf}
%\end{minipage}
%}
%\subfigure[MAE]{
%\begin{minipage}[t]{0.5\textwidth}
%\centering
%\includegraphics[width=8cm]{image/window_size_mae.pdf}
%\end{minipage}
%}
%\caption{\label{fig_reg}Dimension of latent factors.}
%\end{figure}

 %on Douban Movie dataset. As shows in Fig. \ref{fig_reg}, when the values of $\alpha$ and $\beta$ are both around 1, our model has the best performance. When the values of $\alpha$ and $\beta$ are quite large or small, the result is not good, especially in the case of small $\alpha$ and $\beta$. It shows that the latent features of users and items leant from HIN embedding both play an important role for better performances. However, too much weights on latent features of users and items also hurt the performances.

\section{Conclusion \label{sec-con}}
In this paper, we proposed a novel heterogeneous information network embedding based approach (\ie HERec) to effectively utilizing auxiliary information in HINs for recommendation. We designed a new random walk strategy based on meta-paths to derive more meaningful node sequences for network embedding. Since embeddings based on different meta-paths contain different semantic, the learned embeddings were further integrated into an extended matrix factorization model using a set of fusion functions. Finally, the extended matrix factorization model together with fusion functions were jointly optimized for the rating prediction task. HERec aimed to learn useful information representations from HINs guided by the specific recommendation task, which distinguished the proposed approach from existing HIN based recommendation methods. Extensive experiments on three real datasets demonstrated the effectiveness of HERec. We also verified the ability of HERec to alleviate cold-start problem and examine the impact of meta-paths on performance.

As future work, we will investigate into how to apply deep learning methods (\eg convolutional neural networks, auto encoder)
to better fuse the embeddings of multiple meta-paths. In addition, we only use the meta-paths which have the same starting and ending types to effectively extract network structure features in this work. Therefore, it is interesting and natural to extend the proposed model to learn the embeddings of any nodes with arbitrary meta-paths. As a major issue of recommender systems, we will also consider how to enhance the explainablity of the recommendation method based on the semantics of meta-paths.
%As part of future work, we are interested in applying deep learning methods (\emph{e.g.} convolutional neural networks and auto encoders) to better fuse embeddings of various meta-paths. In addition, the proposed HIN embedding can also be employed for other applications.

\bibliographystyle{IEEEtran}
\bibliography{reference}

\begin{IEEEbiography}{Chuan~Shi}
received the B.S. degree from the Jilin University in 2001, the M.S. degree from the Wuhan University in 2004, and Ph.D. degree from the ICT of Chinese Academic of Sciences in 2007. He joined the Beijing University of Posts and Telecommunications as a lecturer in 2007, and is a professor and deputy director of Beijing Key Lab of Intelligent Telecommunications Software and Multimedia at present. His research interests are in data mining, machine learning, and evolutionary computing. He has published more than 40 papers in refereed journals and conferences.
\end{IEEEbiography}

\begin{IEEEbiography}{Binbin~Hu}
received the B.S. degree from Beijing University of Posts and Telecommunications in 2016. He is currently a master student in BUPT. His research interests are in data mining and machine learning.
\end{IEEEbiography}

\begin{IEEEbiography}{Wayne~Xin~Zhao}
is currently an assistant professor at the School of Information, Renmin University of China. He received the Ph.D. degree from Peking University in 2014. His research interests are web text mining and natural language processing. He has published several referred papers in international conferences journals such as ACL, EMNLP, COLING, ECIR, CIKM, SIGIR, SIGKDD, AAAI, ACM TOIS, ACM TKDD, ACM TIST, IEEE TKDE, KAIS and WWWJ.
\end{IEEEbiography}

\begin{IEEEbiography}{Philip~S.~Yu}
is a Distinguished Professor in Computer Science at the University of Illinois at Chicago and also holds the Wexler Chair in Information Technology. Dr. Yu spent most of his career at IBM, where he was manager of the Software Tools and Techniques group at the Watson Research Center. His research inter- est is on big data, including data mining, data stream, database and privacy. He has published more than 920 papers in refereed journals and conferences. He holds or has applied for more than 250 US patents. Dr. Yu is a Fellow of the ACM and the IEEE. He is the Editor-in-Chief of ACM Transactions on Knowledge Discovery from Data. He is on the steering committee of the IEEE Conference on Data Mining and ACM Conference on Information and Knowledge Management and was a member of the IEEE Data Engineering steer- ing committee. He was the Editor-in-Chief of IEEE Transactions on Knowledge and Data Engineering (2001-2004). Dr. Yu received the B.S. Degree in E.E. from National Taiwan University, the M.S. and Ph.D. degrees in E.E. from Stanford University, and the M.B.A. degree from New York University.
\end{IEEEbiography}

\end{document}